\documentclass[3p,amsmath,amssymb]{elsarticle}

\usepackage{amsmath}
\usepackage{graphicx}
\usepackage{latexsym}
\usepackage{amsmath,amssymb}
\usepackage{bm} 
\usepackage{xcolor}
\usepackage{stackrel}
\usepackage{accents}
\usepackage{braket}
\usepackage[title]{appendix}

\journal{Annals of Physics}

\numberwithin{equation}{section}

\DeclareMathOperator{\sgn}{sgn}

\newcommand{\numb}{\addtocounter{equation}{1}\tag{\theequation}}

\DeclareMathOperator{\Tr}{Tr}

\DeclareMathOperator{\Li}{Li}

\newcommand{\vex}[1]{\bm{\mathrm{#1}}}
\newcommand{\msf}[1]{\mathsf{#1}}

\newcommand{\tim}{\mathrm{T}}
\newcommand{\atim}{\bar{\mathrm{T}}}

\newcommand{\dd}{\mathcal{D}}
\newcommand{\D}{\mathcal{D}}
\newcommand{\T}{\mathsf{T}}
\newcommand{\intl}[1]{\int\limits_{#1}}

\newcommand{\kb}{\vex{k}}

\newcommand{\qb}{\vex{q}}

\newcommand{\rb}{\vex{r}}
\newcommand{\xb}{\vex{r}}
\newcommand{\ww}{\omega}

\newcommand{\e}{\varepsilon}

\newcommand{\kf}{k_{\msf{F}}}
\newcommand{\vf}{v_{\msf{F}}}

\newcommand{\Tf}{T_{\msf{F}}}
\newcommand{\Ef}{E_{\msf{F}}}

\newcommand{\phicl}{\phi_{\mathsf{cl}}}
\newcommand{\phiq}{\phi_{\mathsf{q}}}
\newcommand{\mf}{\hat{M}_F}
\newcommand{\htau}{\hat{\tau}}

\newcommand{\im}{\operatorname{Im}}
\newcommand{\re}{\operatorname{Re}}

\begin{document}
	
\sloppy
	
	\begin{frontmatter}

		\title{Know the enemy: 2D Fermi liquids}
		\author[UMD]{Sankar Das Sarma}
		\author[UMD]{Yunxiang Liao}
		\address[UMD]{Condensed Matter Theory Center and Joint Quantum Institute,
			\\Department of Physics, University of Maryland, College Park, MD 20742, USA}

		\date{\today}
		\begin{abstract}
			We describe an analytical theory investigating the regime of validity of the Fermi liquid theory in interacting, via the long-range Coulomb coupling,  two-dimensional Fermi systems comparing it with the corresponding 3D systems.  We find that the 2D Fermi liquid theory and 2D quasiparticles are robust up to high energies and temperatures of the order of Fermi energy above the Fermi surface, very similar to the corresponding three-dimensional situation.  We calculate the phase diagram in the frequency-temperature space separating the collisionless ballistic regime and the collision-dominated hydrodynamic regime for 2D and 3D interacting electron systems.  We also provide the temperature corrections up to third order for the renormalized effective mass, and comment on the validity of 2D Wiedemann-Franz law and 2D Kadawoki-Woods relation.
		\end{abstract}


\end{frontmatter}

\tableofcontents

\section{Introduction}

In a famous talk at the 1989 Kathmandu Summer School, Phil Anderson questioned the validity of the Fermi liquid theory and the applicability of the quasiparticle concept to high-temperature cuprate superconductors specifically and to 2D interacting electron systems generally~\cite{Anderson}.   One of his lectures, which was also quoted in his published lecture note, has the memorable slogan ``\textit{Know the enemy}'', alluding specifically to the Fermi liquid theory as the enemy.  Anderson was the first person vehemently and tirelessly pushing the idea that the 2D physics of cuprate superconductors is beyond the Fermi liquid-BCS paradigm, and represents new physics, which is now commonly referred to as non-Fermi liquids, a terminology virtually unknown in 1989.

The basic challenges Anderson posed were simple:  (1) Is it possible that interactions destroy the 2D Fermi surface just as they do in one dimension leading to a Luttinger liquid?  (2) Is it possible that cuprates represent a new emergent form of superconductivity which simply cannot be explained and understood using the highly successful Fermi liquid-BCS formalism of electron pairing around the Fermi surface leading to a superconducting instability of Copper pairs condensing into a BCS ground state?

Amazingly, there is still no answer to the second question even after 30 years and many thousands of theoretical papers as there is no consensus on the accepted theory of cuprate superconductivity.  In fact, even the precise mechanism  for cuprate superconductivity is still actively debated in the theoretical community, and Anderson himself worked on developing an appropriate theory for the cuprates for the rest of his life.  It is, however, a great testimonial to Anderson's early insight that most theorists working on cuprate superconductivity accept that its explanation most likely lies outside the standard Fermi liquid-BCS paradigm. Our current work is on the continuum 2D interacting Coulombic electron system whereas most work on cuprates uses lattice models with short-range interactions.

But the answer to Anderson's first question is definitively known.  Two dimensional interacting Fermi systems are Fermi liquids similar to 3D interacting Fermi systems, and not non-Fermi liquids like interacting one dimensional Luttinger liquids.  In fact, Anderson's trenchant questioning of the nature of interacting 2D electron systems (``\textit{Know the enemy}'') led to many theoretical developments establishing conclusively, and in fact, even rigorously, that interacting 2D electron systems are indeed normal Fermi liquids with well-defined renormalized Fermi surfaces very much like 3D normal metals and normal He-3~\cite{feldman-1,feldman-2,feldman-3}.   The basic idea, which is now established with reasonable mathematical rigor, is rather simple and essentially the same as it is for 3D interacting Fermi systems.  The imaginary part of the interacting 2D self-energy at energy $E$ goes as $(E-\Ef)^2$ up to some logarithmic corrections, with $\Ef$ being the Fermi energy, and this holds to all orders in perturbation theory.  Thus, the single-particle spectral function is a delta function at the Fermi surface, guaranteeing the existence of an interacting Fermi surface and low-energy quasiparticles with one to one correspondence to the noninteracting Fermi gas.  This behavior is similar to 3D systems within logarithmic accuracy, and thus 2D systems are similar to 3D systems.  This is very different from interacting 1D systems, which within the same perturbation theory  gives an imaginary self-energy going as $(E-E_F) (\ln (E- E_F))^{1/2}$~\cite{Hu}.  Although a perturbation theory is neither meaningful nor valid for a 1D system, it is clear that the Fermi surface cannot exist in 1D already based on this simple perturbative argument.  On the other hand, the perturbative analysis applies in 2D and shows that Fermi surface exists in the 2D interacting systems just as in 3D. 

In the current work, we extend Anderson's first question, trying to understand `the enemy' better.  The question we ask is the extent to which the Fermi liquid theory applies in 2D electron liquids in the presence of long-range Coulomb interactions.  In particular, how far in energy from the Fermi surface and how high in temperature can we go and still find well-defined quasiparticles in 2D interacting systems?  What is the regime of applicability of the concept of 2D Fermi liquids?  We answer these questions analytically, both in 2D and 3D comparing the two situations, using a many-body perturbation theory which is exact for the Coulomb-interacting system in the high-density limit.  

We also calculate the temperature dependent effective mass renormalization, comment on the 2D Wiedemann-Franz law and Kadowaki-Woods relation, and estimate the hydrodynamic regime in Fermi systems interacting through long-range Coulomb interactions.

Our work is completely analytical involving expansions in inverse density, energy, and temperature on an equal footing.  The theory itself uses the well-established leading-order dynamical screening or random phase approximation (RPA) for the self-energy, which is exact in the high-density limit.

\section{Theory}\label{sec:theory}

In this section, we derive the self-energy for an electron system with long-range Coulomb interactions in both 2D and 3D~\cite{AGD,fetter,Galitskii1958,PRB}. In particular, we provide the analytical expressions for both the real and imaginary parts of the on-shell self-energy, for arbitrary energy-to-temperature ratio $\e/T$. Here, $\varepsilon(=E -\Ef)$ is the quasiparticle energy measured from the Fermi surface and $T$ is the temperature. The result is valid to the leading order in $r_s$ and to several orders in $\e/\Ef$ and $T/\Tf$. Here, $r_s$ is the standard dimensionless Coulomb coupling parameter (the ratio of the interparticle separation to the effective Bohr radius), and $\Tf$ is the noninteracting  Fermi temperature.  As usual, $r_s$ is the many body perturbation parameter (or the density-dependent effective fine structure constant) for the theory which is strictly valid only for $r_s\ll1$, but in practice works well empirically for metallic electron densities $(r_s \sim 3-6)$~\cite{Rice}. In Sec.~\ref{sec:formula}, we start with reviewing the derivation of general self-energy formulas which are expressed as integrals involving RPA dynamically screened interaction and are used to extract the explicit self-energy expressions. Sec.~\ref{sec:2Dresult} is devoted to the detailed calculation of the 2D electron self-energy, which is presented before in Ref.~\cite{PRB} and reviewed here for completeness. In Sec.~\ref{sec:3Dresult}, we provide the analytical expressions for the 3D electron self-energy when energy $\e$ and temperature $T$ are arbitrary with respect to each other, which, to the best of our knowledge,  are new for long-range Coulomb interactions. 
See Refs.~\cite{Chubukov2003,Chubukov2004,Chubukov2012} for the results of  self-energy for the case of short-range interactions in both 2D and 3D. We also present the subleading terms in $\e/\Ef$ ($T/\Tf$) for the 3D self-energy in the $\e \gg T$ ($T\gg \e$) limit.

\subsection{General formulas for electron self-energy}\label{sec:formula}

\subsubsection{Keldysh approach to interacting electrons}

In the framework of Keldysh formalism (see Ref.~\cite{Kamenev} for a review), we now rederive the general formulas for self-energy which are then used to obtain the analytic expressions presented in Secs.~\ref{sec:2Dresult} and \ref{sec:3Dresult}. An alternative derivation employing Matsubara technique (see for example Refs.~\cite{AGD,fetter}) will be reviewed in Appendix~\ref{sec:matsubara}.

We start with the partition function $Z$ of a $d-$dimensional electron system with Coulomb interactions on a Keldysh contour which runs from $t=-\infty$ to $t=\infty$ and back to $t=-\infty$. The system is assumed to be in the thermal equilibrium and noninteracting at the distant past $t=-\infty$, after which the interactions are then adiabatically switched on. In terms of coherent state functional integral, the partition function can be expressed as
\begin{align}\label{eq:Z}
\begin{aligned}
	Z
	=&
	\int 
	\dd \left( \bar{\psi}, \psi \right) 
	\,
	\exp
	\left( 
	iS_0+iS_{\msf{int}}
	\right),
\end{aligned}
\end{align}
where $S_0$ and $S_{\msf{int}}$, which stand for the free and interacting parts of the action, respectively, are given by
\begin{align}\label{eq:Z2}
\begin{aligned}
	S_0
	=\,&
	\int_{-\infty}^{\infty} dt 
	\int_{-\infty}^{\infty} dt'
	\int d^d \rb \int d^d \rb'
	\bar{\psi}(\vex{r},t)
	\;
	\hat{G}_0^{-1}(\vex{r}-\vex{r}',t-t')	
	\;
	\psi(\vex{r'},t'),
	\\
	S_{\msf{int}}
	=\,&
	-
	{\frac{1}{2}}
	\,
	\sum_{a = \pm}
	\zeta_a
	\int_{-\infty}^{\infty} dt 
	\int d^d \rb \int d^d \rb'\,
	\bar{\psi}_{\sigma}^{a}(\xb,t) \bar{\psi}_{\sigma'}^{a}(\xb',t)
	V(\xb-\xb')
	\psi_{\sigma'}^{a}(\xb',t)\psi_{\sigma}^{a}(\xb,t).
\end{aligned}
\end{align}
Here $\psi^{+}$($\bar{\psi}^{+}$) and $\psi^{-}$($\bar{\psi}^{-}$)  represent Grassmann fields residue on the forward and backward paths of the Keldysh contour, respectively. They are also labeled by a spin index $\sigma= \uparrow, \downarrow$. The overall sign factor $\zeta_a$ takes the value of $1$ ($-1$) for $a=+$ ($-$).
$V(\rb)=e^2/r$ is the bare Coulomb interaction potential and its Fourier transform is $V(\qb)=2\pi e^2/q$ ($V(\qb)=4\pi e^2/q^2$) in 2D (3D). In the present paper, we work in the units of $\hbar=k_{\msf{B}}=1$, and use $\Ef$ and $T_F$ interchangeably.
$G_0$ is the noninteracting contour ordered electron Green's function, and has the following structure in Keldysh space:
\begin{align}
\begin{aligned}
\hat{G}_0(\vex{r}-\vex{r}',t-t')	
=
-i\braket{\psi(\rb,t) \bar{\psi}(\rb',t')}_0
=
\begin{bmatrix}
G_0^{(\tim)}(\vex{r}-\vex{r}',t-t')	
&
G_0^{(<)}(\vex{r}-\vex{r}',t-t')	
\medskip
\\
G_0^{(>)}(\vex{r}-\vex{r}',t-t')	
&
G_0^{(\atim)}(\vex{r}-\vex{r}',t-t')	
\end{bmatrix},
\end{aligned}	
\end{align}
with $G_0^{(\tim)}$, $G_0^{(\atim)}$, $G_0^{(<)}$ and $G_0^{(>)}$ being the time-ordered, anti-time-ordered, lesser, and greater noninteracting electron Green's function, respectively. 
The angular bracket with subscript $0$ indicates functional integration over Grassmann fields $\psi$ and $\bar{\psi}$ with weight $e^{i S_0}$ {(Eq.~\ref{eq:Z2}).

With the help of a real bosonic field $\phi$, we can perform a Hubbard-Stratonovich transformation to decouple the interactions
\begin{align}\label{eq:HS}
\begin{aligned}
	&
	e^{iS_{\msf{int}}}
	=
	\int 
	\D \phi
	\exp
	\left[
	\frac{i}{2} \sum_{a=\pm }
	\zeta_a
	\intl{\qb,\ww}
	\phi_a(\qb,\ww) V^{-1}(\qb)  \phi_a(-\qb,-\ww)
	-i
	\sum_{a=\pm}
	\zeta_a
	\intl{\e,\kb}\intl{\ww,\qb}
	\phi^{a}(\qb,\ww) 
	\bar{\psi}^{a}_{\sigma}(\kb+\qb,\e+\ww)
	\psi^{a}_{\sigma}(\kb,\e) 
	\right].
\end{aligned}
\end{align}
Here we have used the short-hand notations $\int_{\kb} \equiv \int d^d \kb/(2\pi)^d$ and $\int_{\ww} \equiv \int_{-\infty}^{\infty} d\ww/(2\pi)$.
Introducing the classical and quantum components of the bosonic field $\phi$ :
\begin{align}
	\phicl
	=
	\left( \phi_++\phi_-\right) /\sqrt{2},
	\qquad
	\phiq
	=
	\left( \phi_+-\phi_-\right) /\sqrt{2},
\end{align}
we rewrite Eq.~\ref{eq:HS} as
\begin{align}\label{eq:Sint}
\begin{aligned}
	e^{ i S_{\msf{int}} }
	=\,
	\int 
	\dd \phi
	\,
	\exp
	&
	\left[
	i 
	\intl{\qb,\ww}
	\phicl(\qb,\ww)
	V^{-1}(\qb)
	\phiq(-\qb,-\ww) 
	-
	\frac{i}{\sqrt{2}}
	\intl{\kb,\e, \qb,\ww}	
	\phicl(\qb,\ww) 
	\,
	\bar{\psi}(\kb+\qb,\e+\ww) \htau^3 \psi(\kb,\e) 
	\right. 
	\\
	&\left. 
	-
	\frac{i}{\sqrt{2}}
	\intl{\kb,\e, \qb,\ww}	
	\phiq(\qb,\ww) 
	\,
	\bar{\psi}(\kb+\qb,\e+\ww) \, \psi(\kb,\e) 
	\right],
\end{aligned}
\end{align}
where $\hat{\tau}^{i}$ denotes the Pauli matrix in the Keldysh space.

It is usually convenient to apply the Keldysh rotation to the fermionic fields $\psi$ and $\bar{\psi}$:
\begin{align}\label{CoV1}
\begin{aligned}
	\psi(\kb,\e) 
	\rightarrow \, 
	 \frac{\htau^3 +  \htau^1}{\sqrt{2}} \psi(\kb,\e),
	\qquad
	\bar{\psi}(\kb,\e) 
	\rightarrow \,
	\bar{\psi}(\kb,\e) \,\frac{\hat{1} - i \htau^2}{\sqrt{2}},
\end{aligned}
\end{align}
after which the bare electron Green's function $\hat{G}_0$ assumes the following form in the Keldysh space
\begin{align}
\begin{aligned}
\hat{G}_0(\kb,\e)	
=\,
\begin{bmatrix}
G_0^{(R)}(\kb,\e)	 	& G_0^{(K)}(\kb,\e)	
\\
0 	& G_0^{(A)}  (\kb,\e)
\end{bmatrix}.
\end{aligned}
\end{align} 
In the present paper, we use superscripts $``R"$, $``A"$ and $``K"$ to denote the retarded, advanced and Keldysh components respectively.
The three components of the bare electron Green's function $\hat{G}_0(\kb,\e)$  are connected by the fluctuation-dissipation theorem (FDT):
\begin{align}\label{eq:FDT-G}
G_0^{(K)}(\kb,\e)=\left[ G_0^{(R)}(\kb,\e)-G_0^{(A)}(\kb,\e)\right] \tanh\left( \frac{ \e}{2T}\right) ,
\end{align}
and are given by, respectively,
\begin{align}	\label{eq:GR}
\begin{aligned}
		&	G_0^{(R/A)}(\kb,\e) = \left( \e - \xi_{\kb} \pm i \eta  \right)^{-1}\!\!\!\!\!\!\!,
		\\
		&	 G_0^{(K)}(\kb,\e) 	=-2\pi i\delta \left( \e - \xi_{\kb} \right) \tanh \left( \frac{ \e}{2T} \right) .
\end{aligned}	
\end{align}		
Here $\eta$ is a positive infinitesimal, and $\xi_{\kb}\equiv k^2/2m-\mu$, with $\mu$ being the chemical potential and $m$ the bare mass. We note that $G_0^{(R/A)}(\kb,\e)$ contains information about the spectrum only, while $G_0^{(K)}(\kb,\e)$ knows also about the distribution function.

We then apply another transformation to the fermionic fields $\psi$ and $\bar{\psi}$ 
\begin{align}\label{CoV2}
\begin{aligned}
	\psi(\kb,\e) 
	\rightarrow \, 
	\mf(\e) \, \psi(\kb,\e),
	\qquad
	\bar{\psi}(\kb,\e) 
	\rightarrow \,
	\bar{\psi}(\kb,\e) \, \mf(\e),
\end{aligned}
\end{align}
where $\mf(\e)$ is distribution function dependent and is defined as, in Keldysh space,
\begin{align}\label{eq:MF}
\begin{aligned}
	\mf(\e)
	=
	\begin{bmatrix}
	1  	& \tanh \left( \e/2 T \right) 
	\\
	0 	& -1  
	\end{bmatrix}.
\end{aligned}
\end{align} 
The advantage of this transformation is that the noninteracting electron Green's function becomes diagonal and acquires the form
\begin{align}\label{eq:g0}
\begin{aligned}
\hat{G}_0(\kb,\e)	
=\,
\begin{bmatrix}
G_0^{(R)}(\kb,\e)	 &0
\\
0 	& G_0^{(A)}  (\kb,\e)
\end{bmatrix}.
\end{aligned}
\end{align}
After this transformation, the partition function can now be expressed as
\begin{align}\label{eq:Zrot}
\begin{aligned}
&Z
=\,
\int 
\dd \left( \bar{\psi},\psi \right) 
\dd \phi
\,
\exp
\left( i S_{\psi}+i S_{\phi}+iS_c \right),
\\
&S_{\phi}
=\,
\intl{\qb,\ww}
\phicl(\qb,\ww) V^{-1}(q)\phiq(-\qb,-\ww),
\\
&S_{\psi}
=\,
\intl{\kb,\e}
\bar{\psi}(\kb,\e)
\left( \e - \xi_{\kb} + i \eta \htau^3  \right) 
\psi(\kb,\e),
\\
&S_c
=\,
-\frac{1}{\sqrt{2}}
\intl{\kb,\kb',\e,\e'}
\phicl(\kb-\kb', \e - \e')
\bar{\psi}(\kb,\e) 
\mf(\e)	 \mf(\e') 
\psi(\kb',\e')\,
\\
&\qquad
-\frac{1}{\sqrt{2}}
\intl{\kb,\kb',\e,\e'}
\phiq(\kb-\kb', \e - \e')
\bar{\psi}(\kb,\e) 
\mf(\e) \htau^1 \mf(\e')
\psi(\kb',\e').
\end{aligned}
\end{align}

\subsubsection{RPA dynamically screened interaction}

To proceed, we first integrate out the fermionic fields $\psi$ and $\bar{\psi}$ to get an effective theory of the bosonic field $\phi$:
\begin{align}
\begin{aligned}
	&Z=
	\int 
	\dd \phi\,
	\exp
	\left( 
	 i S_{\phi}
	 +
	 \ln \braket{\exp (iS_c)}_{\psi}
	 \right).
\end{aligned}
\end{align}
Here the angular bracket with a subscript $\psi$ is used to indicate the functional averaging over the fermionic fields $\psi$ and $\bar{\psi}$ with the weight $\exp \left( i S_{\psi} \right) $, so $\braket{\exp (iS_c)}_{\psi}$ represents
\begin{align}
\begin{aligned}
	&\braket{\exp (iS_c)}_{\psi}
	= \,
	\int \dd \left( \bar{\psi},\psi \right)
	\exp \left( i S_{\psi} +iS_c \right).
\end{aligned}
\end{align}
Within the framework of the RPA approximation, we approximate $\braket{\exp (iS_c)}_{\psi}$ by the leading order term in the cumulant expansion:
\begin{align}\label{eq:Sc2}
\begin{aligned}
	&
	\ln \left\langle \exp (iS_c) \right\rangle_{\psi}
	\approx
	 \left\langle \frac{1}{2}(iS_c)^2 \right\rangle _{\psi}.
\end{aligned}
\end{align}
The effective action $\left\langle \frac{1}{2}(iS_c)^2 \right\rangle _{\psi} $ can be expressed in terms of the polarization operator $\hat{\Pi}$ which can be considered as the self-energy of the bosonic field $\phi=(\phi_{\mathsf{cl}},\phi_{\mathsf{q}})^{T}$:
\begin{align}\label{eq:SC2}
\begin{aligned}
	\left\langle \frac{1}{2}(iS_c)^2 \right\rangle _{\psi} 
	=
	-\frac{i}{2} &
	\int_{\qb,\ww}
	\phi^{\T}(-\qb,-\ww) 
	\hat{\Pi}(\qb,\ww)
	\phi(\qb,\ww),
	\\
	\Pi^{ab}(\qb,\ww)
	=\,
	-i
	\intl{\kb,\e}&
	\Tr
	\left\lbrace 
	\left[ 
	\frac{1+\zeta_a}{2}
	+
	\frac{1-\zeta_a}{2}
	\hat{\tau}^1
	\right] 
	\mf (\e+\ww) 
	\hat{G}_0(\kb+\qb,\e+\ww) 
	\mf (\e+\ww)
	\right. 
	\\
	&
	\qquad \left. 
	\times
	\left[ 
	\frac{1+\zeta_b}{2}
	+
	\frac{1-\zeta_b}{2}
	\hat{\tau}^1
	\right] 
	\mf (\e) 
	\hat{G}_0(\kb,\e) 
	\mf (\e)
	\right\rbrace.
\end{aligned}
\end{align}

 Utilizing the causality relation
$
\int_{\kb,\e}
	G_0^{(R)}(\kb+\qb,\e+\ww)G_0^{(R)}(\kb,\e)=0,
$
and inserting the explicit expression for $\mf(\e)$ (Eq.~\ref{eq:MF}) into Eq.~\ref{eq:SC2}, 
one can verify that the polarization operator $\hat{\Pi}$ acquires the following triangular structure in the Keldysh space:	
\begin{align}\label{eq:Pi}
\begin{aligned}
&\hat{\Pi}(\qb,\ww)
=\,
\begin{bmatrix}
0& \Pi^{(A)}(\qb,\ww)
\\
\Pi^{(R)}(\qb,\ww)& \Pi^{(K)}(\qb,\ww)
\end{bmatrix},
\end{aligned}
\end{align}
and its components are connected by the FDT for bosons
\begin{align}\label{eq:Pi3}
\begin{aligned}
&\Pi^{(K)}(\qb,\ww)=\,\left[  \Pi^{(R)}(\qb,\ww) -\Pi^{(A)} (\qb,\ww) \right] 
\coth\left( \frac{ \ww}{2T}\right).
\end{aligned}
\end{align}
One can also show that the retarded (advanced) component $\Pi^{(R/A)} (\qb,\ww)$ is given by
\begin{align}\label{eq:Pi2}
\begin{aligned}
	\Pi^{(R)} (\qb,\ww)
	=\,
	\left( \Pi^{(A)}(\qb,\ww)\right)^*
	=\,&
	-i
	\intl{\kb,\e}
	\left[ 
	G_0^{(R)}(\kb+\qb,\e+\ww)G_0^{(K)}(\kb,\e)
	 +
	G_0^{(K)}(\kb+\qb,\e+\ww)G_0^{(A)}(\kb,\e)
	\right].
\end{aligned}
\end{align}

Employing the explicit form of the bare electron Green's function (Eq.~\ref{eq:GR}), Eq.~\ref{eq:Pi2} can be reduced to the following integral representing the non-interacting polarization function or the irreducible  polarizability (``bare bubble"):
\begin{align}\label{eq:Pi4}
\begin{aligned}
	\Pi^{(R)} (\qb,\ww)
	=\,&
	\intl{\kb}
	\frac{\tanh(\xi_{\kb+\qb}/2T)-\tanh(\xi_{\kb}/2T)}{\ww+\xi_{\kb}- \xi_{\kb+\qb}+i\eta},
\end{aligned}
\end{align}
whose result at zero temperature has been found by Stern~\cite{stern} and Lindhard~\cite{lindhard} in dimensions $d=2$ and $d=3$, respectively.
In terms of the dimensionless parameters $u \equiv \ww/\vf q$ and $x\equiv  q/2\kf$, the zero temperature polarization operator takes the following form in 2D
\begin{align}
&\begin{aligned}\label{eq:pi-2}
	\re \Pi_0^{(R)} (u>0,x)
	=&
	-\nu
	\left\lbrace 
	1
	-
	\frac{\sgn{(x+u)}}{2x}
	\sqrt{\left( x+u\right)^2-1}
	\,
	\Theta \left(   x+u-1\right) 
	\right. 
	\\
	&
	\left. 
	\qquad \quad
	-
	\frac{\sgn{(x-u)}}{2x}
	\sqrt{\left( x-u\right)^2-1}
	\,
	\Theta \left(  \lvert x-u \rvert-1\right) 
	\right\rbrace ,
	\\
	\im \Pi_0^{(R)} (u>0,x)
	=&
	-\nu 
	\left\lbrace 
	-\frac{1}{2x}\sqrt{1-\left( x+u\right)^2}
	\Theta \left( 1- ( x+u ) \right)  
	+\frac{1}{2x}\sqrt{1-\left( x-u\right)^2}
	\Theta \left( 1-\lvert x-u \rvert\right)  
	\right\rbrace,
\end{aligned}
\end{align}
and in 3D it is given by
\begin{align}
&\begin{aligned}\label{eq:pi-3}
&\re \Pi_0^{(R)} (u>0,x)
=
-\nu
\left\lbrace 
\frac{1}{2}
+\frac{1}{8x} \left[ 1-\left( x-u \right)^2\right]  \ln \bigg\lvert \frac{x-u+1}{x-u-1} \bigg\rvert 
+\frac{1}{8x} \left[ 1-\left( x+u \right)^2\right]  \ln \bigg\lvert \frac{x+u+1}{x+u-1} \bigg\rvert
\right\rbrace,  
\\
&\im \Pi_0^{(R)} (u>0,x)
=
-\pi \nu
\left\lbrace 
\frac{u}{2}  \Theta \left( \lvert 1-x\rvert - u \right) 
 +
\frac{1}{8x} \left[ 1-\left( x-u \right)^2\right] \Theta ( 1+x-u) \Theta\left( u-\lvert 1-x \rvert\right) 
\right\rbrace.
\end{aligned}	
\end{align}
Here $\nu$ represents the density of states at Fermi level and is given by $\nu=m/\pi$ and $\nu=m\kf/\pi^2$ in 2D (Eq.~\ref{eq:pi-2}) and 3D (Eq.~\ref{eq:pi-3}), respectively.
We note that Eqs.~\ref{eq:pi-2} and~\ref{eq:pi-3} give the forms of the polarization operator with frequency $\ww\geq 0$. The expressions for $\ww<0$ can be deduced from these equations with the help of $\im \Pi^{(R)} (\qb,\ww)=- \im \Pi^{(R)} (\qb,-\ww)$ and  $\re \Pi^{(R)}  (\qb,\ww)=\re \Pi^{(R)}  (\qb,-\ww)$.

Once the polarization operator is known, it is straightforward to deduce  $\hat{D}(\qb,\ww)$, the interaction dressed Green's function for the bosonic field $\phi$, also known as the RPA dynamically screened interaction potential, from the Dyson equation
\begin{align}\label{eq:D}
\begin{aligned}
	\hat{D}(\qb,\ww)
	\equiv\,&
	-i
	\left\langle 
		\begin{bmatrix}
		\phi_{\msf{cl}}(\qb,\ww) 
		\\
		\phi_{\msf{q}}(\qb,\ww) 
		\end{bmatrix}
		\begin{bmatrix}
		\phi_{\msf{cl}}(-\qb,-\ww) 
		&
		\phi_{\msf{q}}(-\qb,-\ww) 
		\end{bmatrix}
	\right\rangle 
	=
	\left[ \hat{D}_0(\qb,\ww)-\hat{\Pi}(\qb,\ww) \right]^{-1}\!\!\!\!\!\!\!.
\end{aligned}
\end{align}
Here $\hat{D}_0$ is the bare bosonic Green's function arising from the free action $S_{\phi}$ (Eq.~\ref{eq:Zrot}), and takes the form
\begin{align}\label{eq:D0}
\begin{aligned}
	\hat{D}_0(\qb,\ww)
	=
	\begin{bmatrix}
	0 & V(\qb)
	\\
	V(\qb) & 0
	\end{bmatrix}.
\end{aligned}
\end{align}
From Eq.~\ref{eq:D}, one can see that $\hat{D}$ acquires the same structure in Keldysh space as $\Pi^{-1}$:
\begin{align}
\begin{aligned}
\hat{D}(\qb,\ww)
=\,&
\begin{bmatrix}
D^{(K)}(\qb,\ww)& D^{(R)}(\qb,\ww)
\\
D^{(A)}(\qb,\ww)& 0
\end{bmatrix},
\end{aligned}
\end{align}
with components given by
\begin{subequations}
\begin{align}
	&\begin{aligned}\label{eq:Dyson}
	D^{(R)}(\qb,\ww)
	=\,&
	\left[ D^{(A)}(\qb,\ww)\right]^*
	=\,
	\left[ V^{-1}(q)-\Pi^{(R)}(\qb,\ww)\right]^{-1}\!\!\!\!\!\!\!,
	\end{aligned}
	\\
	&\begin{aligned}\label{eq:FDT-D}
	D^{(K)}(\qb,\ww)
	=\,&
	\left[ D^{(R)}(\qb,\ww)-D^{(A)}(\qb,\ww)\right] \coth\left( \frac{\ww}{2T}\right). 
	\end{aligned}
\end{align}
\end{subequations}
A diagrammatic representation of the Dyson equation (Eq.~\ref{eq:Dyson}) for the bosonic field $\phi$ is depicted in Fig.~\ref{fig:D}(a) where the bare (dressed) bosonic propagator $D_0$ ($D$), i.e., the bare (dynamically screened) Coulomb interaction, is represented by the red wavy line with a open (solid) dot in the middle. The black bare bubble corresponds to the polarization operator, with each line representing a bare electron Green's function $G_0$ in  Eq.~\ref{eq:Pi2}.

\begin{figure}[t!]
	\centering
	\includegraphics[width=0.7\linewidth]{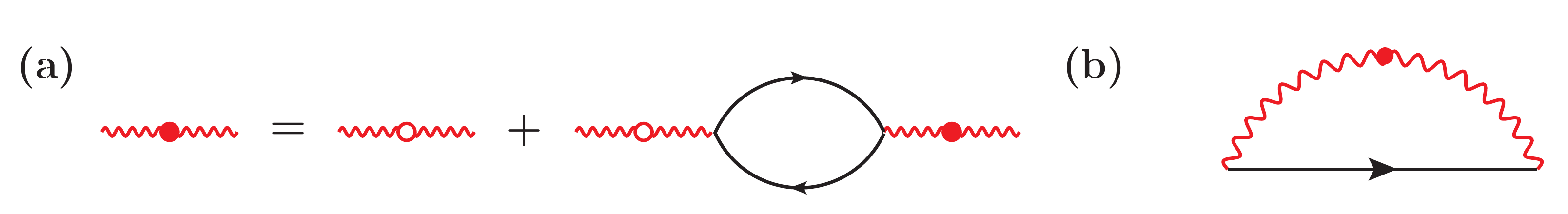}
	\caption{A diagrammatic representation of (a) the Dyson equation for the dynamically screened RPA interaction (Eq.~\ref{eq:Dyson}) and (b) the RPA electron self-energy.
	The dynamically screened (bare) interaction can be considered as the dressed (bare) propagator $D$ ($D_0$) of the bosonic field $\phi$ introduced to decouple the interactions, and is represented diagrammatically by the red wavy line with a solid (open) dot in the middle. 
	The polarization operator $\Pi$ can be considered as the self-energy for bosonic field $\phi$ and is shown by the black bubble where each line represents a bare electron Green's function $G_0$.
	To the leading order in the dynamically screened interaction, the electron self-energy is given by the diagram in panel (b).
 	}
	\label{fig:D}
\end{figure}


\subsubsection{RPA self-energy}

We now return to Eq.~\ref{eq:Zrot},  the partition function before the fermionic fields $\psi$ and $\bar{\psi}$ are integrated out. To the leading order in RPA interaction, the electron self-energy can be obtained from $(iS_c)^2$ the square of the action which couples the fermionic and bosonic fields (Eq.~\ref{eq:Zrot}):
\begin{align}\label{eq:Sigma0}
\begin{aligned}
	-i
	\hat{\Sigma} (\kb,\e) 
	=&
	-\frac{1}{2}
	\intl{\qb,\ww}
	\left\langle 
	\mf(\e) 
	\left[ 
	\phicl(-\qb,-\ww)
	+
	\phiq(-\qb,-\ww)
	\hat{\tau}^1
	\right] 
	\mf(\e+\ww) 
	\psi_{\sigma}  (\kb+\qb,\e+\ww)	
	\right. 
	\\
	&\left. 
	\times
	\bar{\psi}_{\sigma} (\kb+\qb,\e+\ww) 
	\mf(\e+\ww) 
	\left[ 
	\phicl(\qb,\ww)
	+
	\phiq(\qb,\ww)
	\hat{\tau}^1
	\right] 
	\mf(\e) 
	\right\rangle.
\end{aligned}
\end{align}
The angular bracket here represents averaging with respect to the weight:
\begin{align}
\begin{aligned}
	\exp \left( i S_{\psi} +i S_{\phi}+\left\langle \frac{1}{2}(iS_c)^2 \right\rangle _{\psi}  \right) 
	=
	\exp 
	\left(
	i
	 \intl{\kb,\e}
	\bar{\psi}(\kb,\e)
	\hat{G}_0(\kb,\e)
	\psi(\kb,\e) 
	+
	\frac{i}{2}
	\intl{\qb,\ww}
	\phi(\qb,\ww) \hat{D}(\qb,\ww)\phi(-\qb,-\ww)
	\right) .
\end{aligned}
\end{align}
Employing Wick's theorem, we then equate $-i\braket{\phi(\qb,\ww)\phi(-\qb,-\ww)}$ to the RPA dynamically screened interaction $\hat{D}(\qb,\ww)$ and $-i\braket{\psi(\kb,\e)\bar{\psi}(-\kb,-\e)}$ to the bare electron Green's function $\hat{G}_0(\kb,\e)$ in Eq.~\ref{eq:Sigma0}, and obtain
\begin{align}\label{eq:Sig0}
\begin{aligned}
	\hat{\Sigma}(\kb,\e)
	=&
	\frac{i}{2}
	\int_{\qb,\ww}
	D^{(K)}(-\qb,-\ww)
	\mf(\e) 
	\mf(\e+\ww) 
	\hat{G}_0(\kb+\qb,\e+\ww)
	\mf(\e+\ww) 
	\mf(\e) 
	\\
	+\,&
	\frac{i}{2}
	\int_{\qb,\ww}
	D^{(R)}(-\qb,-\ww)
	\mf(\e) 
	\mf(\e+\ww) 
	\hat{G}_0(\kb+\qb,\e+\ww)
	\mf(\e+\ww) 
	\htau^1
	\mf(\e) 
	\\
	+\,&
	\frac{i}{2}
	\int_{\qb,\ww}
	D^{(A)}(-\qb,-\ww)
	\mf(\e) 
	\htau^1
	\mf(\e+\ww) 
	\hat{G}_0(\kb+\qb,\e+\ww)
	\mf(\e+\ww) 
	\mf(\e).
\end{aligned}
\end{align}
$\hat{\Sigma}(\kb,\e)$ is shown diagrammatically by Fig.~\ref{fig:D}(b) where the black solid line and the red wavy line with a closed dot represent, respectively, the bare electron Green's function $G_0$ and the RPA dynamically screened interaction $D$.
Inserting Eqs.~\ref{eq:MF} and~\ref{eq:g0} into the equation above and using the causality relation 
$
\intl{\qb,\ww}
D^{(R)}(\qb,\ww)
G_0^{(R)}(\kb+\qb,\e+\ww)=0,
$
one can prove that $\hat{\Sigma}(\kb,\e)$ is diagonal in the Keldysh space
\begin{align}
\begin{aligned}
	&\hat{\Sigma}(\kb,\e)
	=\,
	\begin{bmatrix}
	\Sigma^{(R)}(\kb,\e) & 0
	\\
	0& 	\Sigma^{(A)}(\kb,\e)
	\end{bmatrix},
\end{aligned}
\end{align}
and its retarded (advanced) component is given by
\begin{align}\label{eq:Sig1}
\begin{aligned}
	&\Sigma^{(R)} (\kb,\e)
	=\,
	\left[ \Sigma^{(A)}(\kb,\e)\right]^*
	=
	\frac{i}{2}
	\intl{\qb,\ww}
	\left\lbrace 
	D^{(K)}(\qb,\ww)G_0^{(R)}(\kb+\qb,\e+\ww)
	+
	D^{(A)}(\qb,\ww)
	G_0^{(K)}(\kb+\qb,\e+\ww)
	\right\rbrace.
\end{aligned}
\end{align}
With the help of Kramers-Kr\"{o}nig relation,
\begin{align}\label{eq:KK}
f^{(R)}(\kb,\e)
=
\int_{-\infty}^{\infty}
\dfrac{d\e'}{\pi}
\dfrac{\im f^{(R)}(\kb,\e')}{\e'-\e-i\eta},
\end{align}
one can rewrite Eq.~\ref{eq:Sig1} as
\begin{align}\label{eq:Sig2}
\begin{aligned}
&\Sigma^{(R)} (\kb,\e)
=\,
-
2
\intl{\qb,\ww,\ww'}
\im D^{(R)}(\qb,\ww)
\im G_0^{(R)}(\kb+\qb,\e+\ww')
\frac{1}{\ww'-\ww-i\eta}
\left[ \coth\left( \frac{\ww}{2T}\right) -\tanh\left( \frac{\e+\ww'}{2T}\right) \right].
\end{aligned}
\end{align}

In two dimensions, after inserting the explicit expression for the bare electron Green's function (Eq.~\ref{eq:GR}) and integrating over the angular direction of the momentum $\qb$, Eq.~\ref{eq:Sig2} reduces to
	\begin{align}\label{eq:Sig3}
	\begin{aligned}
	\Sigma^{(R)} (\kb,\e)
	=\,&
	\frac{m}{\pi k}
	\intl{\ww,\ww'}
	\int_0^{\infty} d q
	\Theta \left( 1-\left| \frac{m\ww'}{kq}\right|\right) 
	\dfrac{ \im D^{(R)}(\qb,\ww)}{
		\sqrt{1-\left( \frac{m\ww'}{kq} \right)^2   } 		
	}
	\\
	&\times
	\dfrac{1}{\ww'-\Delta \e +\frac{q^2}{2m}-\ww-i\eta}
	\left[ \coth\left( \frac{\ww}{2T}\right) -\tanh\left( \frac{\e+\ww'-\Delta \e+\frac{q^2}{2m}}{2T}\right) \right],
	\end{aligned}
	\end{align}
where for simplicity we have defined
\begin{align}
\begin{aligned}\label{eq:de}
		\Delta \e  \equiv \e-\xi_{\kb}.
\end{aligned}
\end{align}
From Eq.~\ref{eq:Sig3}, one then find that the imaginary part of electron self-energy is given by the following integral
\begin{align}
&\begin{aligned}\label{eq:ImSig-0}
	\im \Sigma^{(R)} (\kb,\e)
	=\,
	\frac{m}{4 \pi^2k}
	\int_{-\infty}^{\infty} d\ww
	\left[ \coth\left( \frac{\ww}{2T}\right) -\tanh\left( \frac{\ww+\e}{2T}\right) \right]
	\int_{q_-(\ww)}^{q_+(\ww)}  d q
	\dfrac{ \im D^{(R)}(\qb,\ww)}{
		\sqrt{1 -\left[ \frac{m}{kq}\left( \ww + \Delta \e -\frac{q^2}{2m}\right)\right] ^2   }  },
\end{aligned}
\end{align}
where
\begin{align} \label{eq:qpm}
q_{\pm}(\ww)=
	\left| 
\pm k + \sqrt{k^2 + 2m\left( \ww+\Delta \e \right) }
\right|.
\end{align}
For any $q$ within the regime $q_-(\ww)\leq q \leq q_+(\ww)$, 
$
\left| \frac{m }{kq} \left( \ww+ \Delta \e-\frac{q^2}{2m} \right)  \right| \leq 1
$
is always satisfied.

Applying the Kramers-Kr\"{o}nig relation (Eq.~\ref{eq:KK}),  Eq.~\ref{eq:Sig3} can be rewritten as
\begin{align}\label{eq:Sig-5}
\begin{aligned}
	\Sigma^{(R)} (\kb,\e)
	=\,&
	\frac{m}{4 \pi^2 k}
	\int_0^{\infty} d q
	\int_{-\infty}^{\infty} d\ww'
	\tanh\left( \frac{\e+\ww'-\Delta \e+\frac{q^2}{2m}}{2T}\right)
	\dfrac{ \Theta \left( 1-\left| \frac{m\ww'}{kq}\right|\right) }{
		\sqrt{1-\left( \frac{m\ww'}{kq} \right)^2   } 		
	}
	D^{(A)}(\qb,\ww'-\Delta \e+\frac{q^2}{2m})
	\\
	+&
	\frac{m}{4 \pi^3k}
	\int_{-\infty}^{\infty} d\ww
	\coth\left( \frac{\ww}{2T}\right)
	\int_0^{\infty} d q
	\im D^{(R)}(\qb,\ww)
	\int_{-\frac{kq}{m}}^{\frac{kq}{m}} d\ww'
	\dfrac{1 }{
		\sqrt{1-\left( \frac{m\ww'}{kq} \right)^2   } 		
	}
	\dfrac{1}{\ww'-\Delta \e+\frac{q^2}{2m} -\ww-i\eta}.
\end{aligned}
\end{align}
Taking the real part of this equation, one has
\begin{align}\label{eq:ReSig-0}
\begin{aligned}
		\re \Sigma^{(R)} (\kb,\e)
		=\,&
		\frac{m}{4 \pi^2 k}
		\int_{-\infty}^{\infty} d\ww
		\tanh\left( \frac{\e+\ww}{2T}\right)
		\int_{q_-(\ww)}^{q_+(\ww)} d q
		\dfrac{	\re D^{(R)}(\qb,\ww) }{
			\sqrt{1-\left[  \frac{m }{kq}\left( \ww +\Delta \e-\frac{q^2}{2m}\right) \right] ^2   } 		
		}
		\\
		-&
		\frac{m}{4 \pi^2 k}
		\int_{-\infty}^{\infty} d\ww
		\coth\left( \frac{\ww}{2T}\right)
		\left( \int_0^{q_-(\ww)} d q +\int_{q_+(\ww)}^{\infty} d q \right) 
		\im D^{(R)}(\qb,\ww)
		\frac{	\sgn \left(\ww+ \Delta \e -\frac{q^2}{2m}\right) }{\sqrt{\left[ \frac{m}{kq}\left( \ww+\Delta \e -\frac{q^2}{2m}\right)\right] ^2-1}}.
\end{aligned}
\end{align}

Proceeding in a way analogous to the one outlined above for the 2D self-energy formulas Eqs.~\ref{eq:ImSig-0} and~\ref{eq:ReSig-0}, one can prove that, in 3D, the imaginary and real parts of electron self-energy are given by the following integrals 
\begin{align}\label{eq:Sig6}
\begin{aligned}
	\im \Sigma^{(R)} (\kb,\e)
	=\,&
	\frac{1}{8\pi^2}\frac{m}{k}
	\int_{-\infty}^{\infty} d\ww
	\int_{q_-(\ww)}^{q_+(\ww)} dq 
	q
	\im D^{(R)}(\qb,\ww)
	\left[ \coth\left( \frac{\ww}{2T}\right) -\tanh\left( \frac{\e+\ww}{2T}\right) \right],
	\\
	\re \Sigma^{(R)} (\kb,\e)
	=\,&
	\frac{1}{8\pi^2} \frac{m}{k}
	\int_{-\infty}^{\infty} d\ww
	\int_{q_{-}(\ww)}^{q_{+}(\ww)} dq 
	q \re D^{(R)} (\qb,\ww)	
	\tanh\left( \frac{\ww+\e}{2T}\right)
	\\
	&+
	\frac{1}{8\pi^3}
	\frac{m}{k}
	\int_{0}^{\infty} dq q
	\int_{-\infty}^{\infty} d\ww
	\im D^{(R)}(\qb,\ww)	
	\coth\left( \frac{\ww}{2T}\right) 
	\ln \left[ \left| \dfrac{\ww-\frac{kq}{m}+\Delta\e-\frac{q^2}{2m}}{\ww+\frac{kq}{m}+\Delta\e-\frac{q^2}{2m}}\right| \right].
\end{aligned}	
\end{align}

\subsection{2D electron self-energy}\label{sec:2Dresult}

The self-energy formulas Eqs.~\ref{eq:ImSig-0},~\ref{eq:ReSig-0} and~\ref{eq:Sig6} derived in the previous section are written as two-variables integrals involving the RPA dynamically screened interaction $D^{(R)}(\qb,\ww)$ which is given by another integral for the polarization operator $\Pi^{(R)}(\qb,\ww)$ (Eq.~\ref{eq:Pi4}).
In this section, we will use these formulas to calculate the on-shell ($\e=\xi_{\kb}$) electron self-energy close to the Fermi surface ($k = \kf$) in 2D. We will work in the regime where $r_s^{3/2} \ll \Delta/ \Ef \ll r_s \ll 1$ with $\Delta=\left\lbrace |\e|, T \right\rbrace $, and evaluate the result to the leading order  in $r_s$ and to several orders in $\Delta/(\Ef r_s) $.
Note that the theoretical problem is extremely subtle and challenging since it involves expansions in three distinct `small' parameters: the dimensionless Coulomb coupling $r_s$, temperature $T/T_F$, and energy $\varepsilon/E_F$.    It turns out that the actual small parameters for temperature (energy) expansions are in fact $T/\Tf r_s$ ($\varepsilon/\Ef r_s$), but we allow $T$ and $\varepsilon$ to be comparable in our theory, which considerably complicates the calculations.

Using Eq.~\ref{eq:ImSig-0} and~\ref{eq:ReSig-0}  and setting $\Delta \e=0$, we find that the real and imaginary parts of the on-shell self-energy in 2D are given by 
	\begin{subequations}\label{eq:S}
	\begin{align}
	&\begin{aligned}\label{eq:ImS-0}
	\im \Sigma^{(R)} (\e)
	=\,&
	\int_{0}^{\infty} \frac{d\ww}{2\pi}
	\left[2 \coth\left( \frac{\ww}{2T}\right) 
	-\tanh\left( \frac{\ww+\e}{2T}\right)-\tanh\left( \frac{\ww-\e}{2T}\right)
	\right]
	\im  I(\ww),
	\end{aligned}	
	\\
	&\begin{aligned}\label{eq:ReS-0}
	\re \Sigma^{(R)} (\e)
	=\,
	\int_{0}^{\infty} 
	\frac{d\ww}{2\pi}
	\left[ 
	\tanh\left( \frac{\ww+\e}{2T}\right)
	-
	\tanh\left( \frac{\ww-\e}{2T}\right)
	\right] 
	\re I(\ww),
	\end{aligned}
	\end{align}
	\end{subequations}
	where 
	\begin{align}\label{eq:I}
	I(\ww)
	\equiv
	\frac{m}{2 \pi k}
	\int_{q_-(\ww)}^{q_+(\ww)} d q
	\dfrac{	D^{(R)}(\vex{q},\ww) }{
		\sqrt{1-\left[  \frac{m }{kq}\left( \ww -\frac{q^2}{2m} \right) \right] ^2   } 		
	}.
	\end{align}
Here we have neglected the second integral in Eq.~\ref{eq:ReSig-0} which vanishes to the leading order in $r_s$.

\subsubsection{Momentum integration}

We will now evaluate $I(\ww)$ defined in Eq.~\ref{eq:I} by performing the momentum integration.
For convenience, we introduce
\begin{align}\label{eqn:notation}
\begin{aligned}
	\alpha \equiv \frac{r_s}{\sqrt{2}},
	\quad
	\delta &\equiv \frac{\omega}{4\Ef},
	\quad
	x \equiv \frac{q}{2k_{\msf{F}}},
	\quad
	x_{\pm}(\delta) \equiv \frac{q_{\pm}(\ww)}{2k_{\msf{F}}}.
\end{aligned}
\end{align}
Here $x_{\pm}(\delta)$ is a solution to the equation $(x-\delta/x)^2=1$, and is given by
\begin{align*}
	x_-(\delta)&=\frac{1}{2}\Big|1-\sqrt{1+4\delta}\,\Big|,
	\quad
	x_+(\delta)=\frac{1}{2}\Big(1+\sqrt{1+4\delta}\Big).
\numb
\end{align*}

Using the newly introduced variables, Eq.~\ref{eq:I} can then be rewritten as
\begin{align*}
	I(\delta)
	=&
	\frac{m}{\pi}
	\int_{x_-(\delta)}^{x_+(\delta)}dx\;D^{(R)}(x,\delta)\;
	\Bigg[1-\left( x-\frac{\delta}{x}\right) ^2\Bigg]^{-1/2}
	\!\!\!\!\!
	\\
	=&
	\frac{m}{\pi}
	\int_{x_-(\delta)}^{x_+(\delta)}dx\;D^{(R)}(x,\delta)\;x
	\Big[\left( x_+^2(\delta)-x^2\right) \left( x^2-x_-^2(\delta)\right) \Big]^{-1/2}
	\!\!\!\!\!\!\!,
\numb\label{eqn:int}
\end{align*}
where the retarded RPA interaction $D^{(R)}(x,\delta)$ (Eq.~\ref{eq:Dyson}) is given by
\begin{equation}
	D^{(R)}(x,\delta)= \nu^{-1}
	\dfrac{\left( \frac{x}{\alpha}-\nu^{-1}\re\Pi_0^{(R)}(x,\delta)\right) +i\left( \nu^{-1}\im\Pi_0^{(R)}(x,\delta)\right) }
	{\left( \frac{x}{\alpha}-\nu^{-1}\re\Pi_0^{(R)}(x,\delta)\right)^2+\left( \nu^{-1}\im\Pi_0^{(R)}(x,\delta)\right)^2}.
\label{eqn:vr}
\end{equation}
Here we have approximated the polarization bubble by the zero-temperature result $\Pi_0$ (Eq.~\ref{eq:pi-2}) whose real and imaginary parts, in terms of the new variables, take the forms
\begin{subequations}\label{eq:Pi0}
	\begin{align}
	&\begin{aligned}\label{eqn:pire}
	\nu^{-1}\re\Pi_0^{(R)}(x,\delta)
	=&
	-1
	+\frac{1}{2x^2}\sgn\bigg(1-\frac{\delta}{x^2}\bigg)
	\re\sqrt{-\left( x_+^2(\delta)-x^2\right) \left( x^2-x_-^2(\delta)\right) }
	\\
	&
	+\frac{1}{2x^2}\sgn\bigg(1+\frac{\delta}{x^2}\bigg)
	\re\sqrt{4\delta x^2-\left( x_+^2(\delta)-x^2\right) \left( x^2-x_-^2(\delta)\right) },
	\end{aligned}
	\\
	&\begin{aligned}\label{eqn:piim}
	\nu^{-1} \im\Pi_0^{(R)}(x,\delta)
	=&
	-
	\frac{1}{2x^2} \re\sqrt{\left( x_+^2(\delta)-x^2\right) \left( x^2-x_-^2(\delta)\right) }
	+
	\frac{1}{2x^2}\re\sqrt{\left( x_+^2(\delta)-x^2\right) \left( x^2-x_-^2(\delta)\right) -4\delta x^2}.
	\end{aligned}
	\end{align}
\end{subequations}

 It is clear that the main contribution to the integral in Eq.~\ref{eq:S} is from $\ww$ of the order of $|\e|$ or $T$ due to the presence of the thermal factor which involves tanh and coth functions. As mentioned earlier, we consider the regime where $r_s^{3/2} \ll \Delta/ \Ef \ll r_s \ll 1$, with $\Delta=\left\lbrace |\e|, T \right\rbrace $, and therefore will calculate  $I(\delta)$ for $\alpha^{3/2}\ll |\delta| \ll \alpha \ll 1$, to the leading order in $\alpha$ and several orders in $|\delta|/\alpha$.

To proceed, we divide the interval of integration in Eq.~\ref{eqn:int} into three subintervals: $(x_-,l_-)$, $(l_-,l_+)$ and $(l_+,x_+)$.
Here $l_{\pm}$ can take arbitrary value as long as it satisfies $x_- \ll l_- \ll \alpha \ll l_+ \ll x_+$.  
Within these subregions, we then expand the integrand using different small parameters, such as $x/\alpha$, $\delta/x$, $x/x_+$ and $\alpha/x$. We note that these variables are small in some subregions but become comparable to or larger than $1$ in other subregions. In other words, it is not possible to use a single small parameter for the entire integration interval.
This lack of a single unique small parameter in the theory has prevented the regime of $\varepsilon \sim T$, while both being small, being studied in the theoretical literature at all in spite of the long history of electronic many body field theories~\cite{AGD,fetter,Chubukov2004}.

We set the real part of the polarization operator to
\begin{equation}\label{eq:repi0}
	\nu^{-1}\re\Pi_0^{(R)}(x,\delta)=-1,
\end{equation}
which is valid as long as $x$ lies within the region $x\in (a_-,a_+)$. $a_{\pm}$ is defined as $a_-=x_-[1+O(|\delta|^{1-b})]$ and $a_+=x_+[1-O(|\delta|)]$ where $b \in (0, 1)$ is arbitrary.
Furthermore, for $x\in(a_-,l_+)$, 
$\im \Pi_0^{(R)}$ (Eq.~\ref{eqn:piim}) can be approximated as
\begin{align*}\label{eq:impi0}
	\nu^{-1}\im\Pi_0^{(R)}(\sqrt{z^2+x_-^2},\delta)
	=&-
	\frac{zx_+}{2(z^2+x_-^2)}\Big[1-\sqrt{1-\frac{4\delta (z^2+x_-^2)}{z^2x_+^2}}+O\Big(\frac{l_+^2}{x_+^2}\Big)\Big]
	\\=&-
	\frac{zx_+}{2(z^2+x_-^2)}\frac{4\delta (z^2+x_-^2)}{2z^2x_+^2}\Big[1+O\big(|\delta|^b\big)\Big]
	=-
	\frac{\delta}{z}\Big[1+O\big(|\delta|^b\big)\Big],
	\numb
\end{align*}
where $z^2=x^2-x_-^2$.
For $x\in(l_+,a_+)$, all we need to know is that
\begin{equation}
	\nu^{-1}\im\Pi_0^{(R)}(x,\delta)
	=O\Big(|\delta|^{1/2},\frac{\delta}{l_+}\Big).
\end{equation}

We will now show that the contribution to $I(\delta)$ from $x$ outside the region $x\in (a_-,a_+)$ is of higher order in $\delta$ and therefore can be ignored.
Let us first calculate $I(x_-,a_-)$, i.e., the contribution to $I(\delta)$ from $x_-<x<a_-$. Throughout this section, $I(x_1,x_2)$ is used to denote the contribution to $I(\delta)$ from the integration over the interval $x\in (x_1,x_2)$.
Applying a change of variables $z^2=x^2-x_-^2$, one obtains
\begin{align*}
	&I(x_-,a_-)=\int_0^{O(|\delta|^{\frac{3-b}{2}})}dz\;D^{(R)}(\sqrt{z^2+x_-^2},\delta)\,\Big(x_+^2-z^2-x_-^2\Big)^{-1/2}
	=O(|\delta|^{\frac{3-b}{2}}).
	\numb
\end{align*}
Here we have used the fact that $|D^{(R)}(x,\delta)|\le\nu^{-1}\alpha \,O(\alpha^{-1})$ inside the region $x\in (x_-,a_-)$.
Similarly, for the region $a_+<x<x_+$, $|\nu^{-1}\Pi_0^{(R)}(x,\delta)|=O(|\delta|^{1/2})$ which leads to $|D^{(R)}(x,\delta)|=\nu^{-1}O(\alpha)$.  Applying the transformation $u^2=x_+^2-x^2$, we find the contribution to $I(\delta)$ from this region 
\begin{align*}
	&I(a_+,x_+)=\int_0^{O(|\delta|^{1/2})}du\;D^{(R)}(\sqrt{x_+^2-u^2},\delta)\,\Big(x_+^2-u^2-x_-^2\Big)^{-1/2}=\alpha\;O(|\delta|^{1/2}).
	\numb
\end{align*}

We have proved that the contribution from outside the regime $x\in (a_-,a_+)$ can be ignored, and will now evaluate  $\im I(\delta)$ by carrying out the integration in Eq.~\ref{eqn:int} separately for the three subintervals: $(a_-,l_-)$, $(l_-,l_+)$, and $(l_+,a_+)$.
Within the first region  $x\in (a_-,l_-)$, the integrand can be expanded in terms of the small parameter $x/\alpha \ll 1$.
Employing the transformation $z^2=x^2-\delta^2$, and 
inserting Eqs.~\ref{eq:repi0} and~\ref{eq:impi0} into Eq.~\ref{eqn:int}, one obtains
\begin{align*}
	&\im I(a_-,l_-)=
	-
	\int_{O(|\delta|^{\frac{3-b}{2}})}^{l_--\delta^2\!/(2l_-)}dz\;\frac{\delta/z}{(\frac{\sqrt{z^2+\delta^2}}{\alpha}+1)^2+(\delta/z)^2}\\
	&=
	-
	\int_{O(|\delta|^{\frac{3-b}{2}})}^{l_--\delta^2\!/(2l_-)}dz\;\frac{\delta}{z}\bigg(1+\Big(\frac{\delta}{z}\Big)^2\bigg)^{-1}
	\bigg[1-\frac{2\sqrt{z^2+\delta^2}}{\alpha}\bigg(1+\Big(\frac{\delta}{z}\Big)^2\bigg)^{-1}\bigg]\\
	&=
	-
	{\alpha}
	\bigg[\frac{\delta}{\alpha}\ln\left( \frac{l_-}{|\delta|}\right) +4\frac{\delta^2}{\alpha^2}\sgn\delta-2\frac{l_-\delta}{\alpha^2}\bigg].
	\numb
\end{align*}
For $x$ lying inside the region $(l_-,l_+)$, we instead expand in terms of $|\delta|/x\ll 1$ and $x \ll 1$, which leads to
\begin{align*}
	&\im I(l_-,l_+)=
	-
	\int_{l_-}^{l_+}dx\;\frac{\delta\alpha^2}{x(x+\alpha)^2}
	=
	-
	\alpha
	\bigg[-\frac{\delta}{\alpha}+\frac{\delta}{\alpha}\ln\left( \frac{\alpha}{l_-}\right) +2\frac{l_-\delta}{\alpha^2}\bigg].
	\numb
\end{align*}
Finally, the contribution from $x\in (l_+,a_+)$, i.e., $\im I(l_+,a_+)$, is negligible as the integrand is of the order of $\alpha$.
Combining  contributions from all subregions, we find the imaginary part of $I(\delta)$
\begin{equation}
	\im I (\delta)=
	-
	\alpha
	\left\lbrace 
	\frac{\delta}{\alpha}\left[ -\!1+\ln \left( \frac{\alpha}{|\delta|} \right) \right]  
	 +4\frac{\delta^2}{\alpha^2} \sgn\delta
	\right\rbrace .
\end{equation}

We then turn to the calculation of the real part of $I(\delta)$ by considering the three subregions separately as in the case of $\im I(\delta)$.
For $a_-\leq x \leq l_-$, we use $x/\alpha \ll 1$ as the small expansion parameter, and approximate $\re D^{(R)}$ by
\begin{align*}
	&\re D^{(R)}(x,\delta)
	=\nu^{-1}\frac{1+\frac{x}{\alpha}}{(1+\frac{x}{\alpha})^2+\frac{\delta^2}{z^2}+O(|\delta|^{\frac{3b-1}{2}})}
	=\nu^{-1}\frac{1+\frac{x}{\alpha}}{1+\frac{\delta^2}{z^2}}\bigg[1-\frac{\frac{2x}{\alpha}+\frac{x^2}{\alpha^2}}{1+\frac{\delta^2}{z^2}}+\bigg(\frac{\frac{2x}{\alpha}}{1+\frac{\delta^2}{z^2}}\bigg)^2+O\Big(\frac{x^3}{\alpha^3}\Big)\bigg],
	\numb
\end{align*}
where $z^2=x^2-x_-^2$. This leads to
\begin{align*}
	&\re I(a_-,l_-)=
	\nu
	\int_{O(|\delta|^{\frac{3-b}{2}})}^{l_--\delta^2\!/(2l_-)}dz\;\re D^{(R)}(\sqrt{z^2+x_-^2},\delta)\,\Big[1+O(l_-^2)\Big]
	\\
	&=\alpha
	\bigg[
	\frac{z}{\alpha}-\frac{z(5\delta^2+z^2)}{2\alpha^2\sqrt{\delta^2+z^2}}-\frac{\delta}{\alpha}\arctan\left( \frac{z}{\delta}\right) 
	+\frac{5\delta^2}{2\alpha^2}
	\ln \big(z+\sqrt{\delta^2+z^2}\big)
	\bigg]\bigg|_{O(|\delta|^{\frac{3-b}{2}})}^{l_--\delta^2\!/(2l_-)}\\
	&=\alpha
	\bigg[
	-\frac{\pi|\delta|}{2\alpha}+\frac{l_-}{\alpha}
	-\frac{7\delta^2}{4\alpha^2}
	+\frac{5\delta^2}{2\alpha^2}\ln\left( \frac{2l_-}{|\delta|}\right) 
	+\frac{\delta^2}{2\alpha l_-}-\frac{l_-^2}{2\alpha^2}+O\Big(\frac{l_-^3}{\alpha^3},\;\frac{\delta^3}{l_-^3},\;\frac{|\delta|^{\frac{3-b}{2}}}{\alpha},\;|\delta|^{\frac{3b-1}{2}}\Big)\bigg].
	\numb
\label{eqn:reia1l1}
\end{align*}
On the other hand, for the region $x\in(l_-,l_+)$, we expand in terms of $|\delta|/x\ll1$ and $x\ll 1$ instead of $x/\alpha$. Substituting
\begin{equation}
\re D^{(R)}(x,\delta)=\nu^{-1}\frac{1+\frac{x}{\alpha}}{(1+\frac{x}{\alpha})^2+\frac{\delta^2}{x^2}+O(|\delta|^{\frac{3b-1}{2}},\,\frac{\delta^4}{x^4})},
\end{equation}
into Eq.~\ref{eqn:int}, one obtains
\begin{align*}
	&\re I(l_-,l_+)
	=\nu
	\int_{l_-}^{l_+}dx\;\re D^{(R)}(x,\delta)\,\bigg[1+\frac{\delta^2}{2x^2}+O\Big(\frac{\delta^3}{x^3},x\Big)\bigg]\\
	&=
	\alpha
	\bigg[-\frac{l_-}{\alpha}-\frac{5\delta^2}{2\alpha^2}
	+\frac{5\delta^2}{2\alpha^2}\ln\left( \frac{\alpha}{l_-}\right) 
	-\frac{\delta^2}{2\alpha l_-}+\frac{l_-^2}{2\alpha^2}
	+\ln \left( \frac{l_+}{\alpha}\right) +O\Big(\frac{l_-^3}{\alpha^3},\;\frac{\delta^3}{l_-^3},\;\frac{\alpha}{l_+},\;l_+\Big)\bigg].
	\numb
\end{align*}
For the last region $x\in (l_+,a_+)$, $\alpha/x \ll 1$ now becomes the small expansion parameter. Keeping only the leading order term in $\alpha/x \ll 1$, we have
\begin{equation}
	D^{(R)}(x,\delta)=\nu^{-1}\Big[\frac{\alpha}{x}+O\Big(\frac{\alpha^2}{x^2}\Big)\Big],
\end{equation}
which leads to
\begin{align*}
	&\re I(l_+,a_+)=\int_{l_+}^{1+O(\delta)}dx\;\Big[\frac{\alpha}{x}+O\Big(\frac{\alpha^2}{x^2}\Big)\Big](1-x^2)^{-1/2}
	=\alpha \Big[\ln \left( \frac{2}{l_+}\right) +O\Big(\frac{\alpha}{l_+},\;l_+\Big)\Big].
	\numb
\end{align*}
Combining everything, we arrive at 
\begin{align*}
	&\re I (\delta)=\alpha
	\bigg[
	\ln\left( \frac{2}{\alpha}\right) -\frac{\pi|\delta|}{2\alpha}
	+\frac{\delta^2}{\alpha^2}\Big(-\frac{17}{4}+\frac{5}{2}\ln\left( \frac{2\alpha}{|\delta|}\right) \Big)
	+O\Big(\frac{\delta^3}{\alpha^3},\;\alpha,\;|\delta|^{\frac{1-b}{2}},\;|\delta|^{\frac{3b-1}{2}}\Big)\bigg].
	\numb
\end{align*}

\subsubsection{Frequency integration}

The calculation presented in the previous section shows that
\begin{subequations}
\begin{align}
&\begin{aligned}\label{eq:I_1}
\im I(\ww)
=\, &
-
\left\lbrace 
\frac{1}{4}\frac{|\ww|}{\Ef}
\left[ \ln \left( \frac{2\sqrt{2} r_s \Ef}{|\ww|}\right) -1 \right] 
+\frac{1}{2\sqrt{2}r_s} \left( \frac{\ww}{\Ef}\right)^2 \right\rbrace 
\sgn \ww,
\end{aligned}
\\
&\begin{aligned}\label{eq:I_2}
\re I(\ww)
=
\frac{r_s}{\sqrt{2}}
\left[ 
\ln \left( \frac{2\sqrt{2}}{r_s} \right) 
-\frac{\pi}{4\sqrt{2}r_s}   \frac{|\ww|}{\Ef} 
+\frac{5 }{16 r_s^2} \frac{\ww^2}{\Ef^2} \ln \left( \frac{4 \sqrt{2} r_s \Ef}{|\ww|}\right) 
-\frac{17}{32 r_s^2} \frac{\ww^2}{\Ef^2}
\right],
\end{aligned}
\end{align}
\end{subequations}
where we have transformed back to the original variables.
These expressions can be used to extract the electron self-energy on the mass shell by substitution using Eq.~\ref{eq:S} and carrying out the frequency integrations. 

The frequency integrations in Eq.~\ref{eq:S} involve hyperbolic functions $\tanh(x)$ and $\coth(x)$, and are of the form
\begin{align}\label{eq:I12}
\begin{aligned}
I_1(a)
=&
\int_0^{\infty}
dx
f(x) 
\left[ 
2\coth (x)
-\tanh (x+a)
-\tanh (x-a)
\right],
\\
I_2(a)
=&
\int_0^{\infty}
dx
f(x)
\left[ 
\tanh (x+a)
-\tanh (x-a)
\right].
\end{aligned}	
\end{align}
To carry out such integrals, one may take advantage of the following equations which express $\tanh(x)$ and $\coth(x)$ as infinite exponential series:
\begin{align}\label{eq:expser}
\begin{aligned}
&
\tanh (x)
=\,
1+2\sum_{k=1}^{\infty} (-1)^k e^{-2kx},
\qquad
\coth (x)
=\,
1+2\sum_{k=1}^{\infty}  e^{-2kx},
\qquad
x>0.
\end{aligned}
\end{align}
In Appendix.~\ref{sec:integrals}, we use the equations above and provide the analytical results for integrals
of the form Eq.~\ref{eq:I12} for various functions $f(x)$.

Using Eq.~\ref{eq:hy} in Appendix~\ref{sec:integrals},  we obtain the analytical expression for the imaginary part of the 2D self-energy on the mass shell:
	\begin{align}\label{eq:ImSig}
	\begin{aligned}
	\im \Sigma^{(R)}  ( \e, T)
	=&
	-
	\frac{1}{8 \pi }\left( \pi^2+\frac{\e^2}{T^2}\right)
	\frac{T^2}{\Ef}  \ln \left( \frac{\sqrt{2} r_s \Ef}{T}\right) 
	\\
	&
	-
	\left\lbrace 
	-\frac{\pi}{24}
	\left(6- \gamma_E -\ln \left( \frac{2}{\pi^2}\right)  -24\ln \mathrm{A} \right) 
	-\frac{\left(2-\gamma_E-\ln 2 \right)}{8 \pi } \frac{\e^2}{T^2} 
	\right. 
	\\
	&
	\left. 
	\qquad
	+\frac{1}{4 \pi }
	\left[ \partial_s \Li_s (-e^{-\e/T})+ \partial_s \Li_s (-e^{\e/T})\right] \bigg \lvert_{s=2} 
	\right\rbrace 
	\frac{T^2}{\Ef} 
	\\
	&
	-
	\frac{\sqrt{2}}{\pi}
	\left[ 
	\zeta(3)
	-\frac{1}{2}\Li_3 (-e^{\e/T})	
	-\frac{1}{2}\Li_3 (-e^{-\e/T})
	\right]
	\frac{T^3}{r_s \Ef^2}.
	\end{aligned}
	\end{align}
Here $\gamma_E \approx 0.577216$ is the Euler's constant, and $\mathrm{A} \approx 1.28243$ is the Glaisher's constant.
$\Li_s(z)=\sum_{k=1}^{\infty} {z^k}/{k^s}$ stands for the polylogarithm function, and $\zeta(z)=\sum_{k=1}^{\infty} {1}/{k^z}$ represents the Riemann zeta function. We emphasize that this result is valid for arbitrary value of $\e/T$, as long as $|\varepsilon|, T \ll \Ef$ condition is satisfied.
 
 One can derive from Eq.~\ref{eq:ImSig} the asymptotic behaviors of $\im \Sigma^{(R)}$ in the low-energy limit $|\e| \ll T$ as well as the low-temperature limit $T \ll |\e| $.
 With the help of
 \begin{align}
 \begin{aligned}
 \partial_s \Li_s (-1)|_{s=2}
 =\,
 -\frac{\pi^2}{12}\left(\gamma_E+\ln \left( 4\pi\right) -12\ln \mathrm{A} \right),
 \qquad
 \Li_3 (-1)=-\frac{3}{4}\zeta(3),
 \end{aligned}
 \end{align}
we find that, for $|\e |\ll T$, $\im \Sigma^{(R)}$ assumes the form
\begin{align}\label{eq:ImSig-<}
\begin{aligned}
	&\im \Sigma^{(R)} (|\e |\ll T)
	=\,
	-
	\frac{\pi}{8} 
	\frac{ T^2 }{\Ef} 
	\ln \left( \frac{\sqrt{2} r_s \Ef}{T}\right) 
	+
	\frac{\pi}{24}
	\left(6 + \ln \left( 2\pi^3\right)  -36 \ln \mathrm{A} \right) 
	\frac{ T^2}{\Ef}
	- 
	\frac{7\zeta(3)}{2\sqrt{2}\pi} \frac{T^3}{r_s\Ef^2}.
\end{aligned}
\end{align}
Similarly for $T \ll |\e|$, we have
\begin{align}\label{eq:ImSig->}
\begin{aligned}
	&\im \Sigma^{(R)} (|\e| \gg T)
	=\,
	-
	\frac{\e^2 }{8 \pi \Ef}  \ln \left( \frac{\sqrt{2} r_s \Ef}{|\e| }\right) 
	-
	\left( \ln 4 -1\right)
	\frac{\e^2 }{16 \pi \Ef} 
	-
	\frac{1}{6\sqrt{2} \pi}\frac{|\e|^3}{r_s \Ef^2},
\end{aligned}
\end{align}
where we have used 
\begin{align}\label{eq:polylog-lim}
\begin{aligned}
&\lim\limits_{x \rightarrow \infty} 
\Li_s (-e^{x})
=-\frac{x^s}{\Gamma\left( s+1 \right) },
\quad
\lim\limits_{x \rightarrow \infty} 
\partial_s \Li_s (-e^{x})
=-\frac{x^s\ln x}{\Gamma\left( s+1 \right) } +\frac{\Gamma' \left( s+1\right) }{\Gamma^2 \left( s+1 \right)}x^s.
\end{aligned}
\end{align}

In a way analogous to the one that leads to the expression of the imaginary part of self-energy, we evaluate the integration in Eq.~\ref{eq:S} using Eq.~\ref{eq:hy} in Appendix~\ref{sec:integrals},  and derive the real part of the on-shell self-energy for arbitrary $\e/T$ in 2D:
\begin{align}\label{eq:ReSig}
\begin{aligned}
&\re \Sigma^{(R)} (\e, T)
=\,
\frac{r_s}{\sqrt{2} \pi}
\ln \left( \frac{2\sqrt{2}}{r_s} \right) \e
-
\frac{1}{8}	
\frac{T}{\e}
\left[ \Li_2(-e^{-\frac{\e}{T}}) - \Li_2(-e^{\frac{\e}{T}}) \right]
\frac{T \e}{\Ef}
+
\frac{5}{48\sqrt{2} \pi}
\left(  \pi^2 +  \frac{\e^2}{T^2} \right) 
\frac{ T^2 \e}{r_s \Ef^2} 
\ln \left( \frac{ r_s \Ef}{ T }\right)
\\
&
-
\left\lbrace
\frac{1}{96 \sqrt{2} \pi }
\left( 32-10\gamma_E-25\ln 2 \right) 
\left( \frac{\e^2}{T^2} +\pi^2  \right) 
+
\frac{5}{8 \sqrt{2} \pi }
\frac{T}{\e}
\left[ \partial_s \Li_s (-e^{-\frac{\e}{T}}) - \partial_s \Li_s (-e^{\frac{\e}{T}}) \right] \bigg\lvert_{s=3}
\right\rbrace 
\frac{ T^2 \e}{r_s \Ef^2}.
\end{aligned}
\end{align}

In the limit of $|\e|/T\ll 1$, one has
\begin{align}\label{eq:polylog-1}
\begin{aligned}
	&\Li_2(-e^{-\frac{\e}{T}}) - \Li_2(-e^{\frac{\e}{T}}) 
	=
	2\ln 2 \frac{\e}{T} +O\left( (\frac{\e}{T} )^2\right) ,
	\\
	&\left[ \partial_s \Li_s (-e^{-\frac{\e}{T}}) - \partial_s \Li_s (-e^{\frac{\e}{T}}) \right] \bigg\lvert_{s=3}
	=
	2\partial_s \partial_z \Li_s (z)|_{s=3, z=-1} \frac{\e}{T} 
	+O\left( (\frac{\e}{T} )^2\right) ,
\end{aligned}
\end{align}
where $\partial_s \partial_z \Li_s (z)|_{s=3, z=-1}$ can be further simplified to 
$(\pi^2\ln2 +6\zeta'(2))/12$.
Substitution of Eq.~\ref{eq:polylog-1} into Eq.~\ref{eq:ReSig}
leads to the asymptotic expression for the real part of self-energy in the limit  of $|\e| \ll T$, 
\begin{align}\label{eq:ReSig-<}
\begin{aligned}
\re \Sigma^{(R)} (|\e| \ll T)
=\,&
\frac{r_s}{\sqrt{2} \pi}
\ln \left( \frac{2\sqrt{2}}{r_s} \right) 
\e
-
\frac{ \ln 2}{4} 
\frac{T \e}{\Ef}
+
\frac{5\pi}{48\sqrt{2} r_s}
\frac{ T^2 \e }{\Ef^2} 
\ln \left( \frac{r_s \Ef}{ T }\right)
\\
&
+
\left[
 -
\frac{\pi}{96 \sqrt{2} }
\left(32-10\gamma_E-25\ln 2 \right) 
-
\frac{5}{8 \sqrt{2} \pi }
\left( \zeta ' (2) + \frac{\pi^2}{6}\ln 2\right) 
\right] 
\frac{ T^2 \e }{\Ef^2 r_s} .
\end{aligned}
\end{align}
The expression for $\re \Sigma^{(R)}$ in the low-temperature limit $T \ll |\e|$ can also be extracted from Eq.~\ref{eq:ReSig}. With the help of Eq.~\ref{eq:polylog-lim}, we arrive at the result
\begin{align}\label{eq:ReSig->}
\begin{aligned}
&\re \Sigma^{(R)} (|\e| \gg T)
=\,
\frac{r_s}{\sqrt{2} \pi}
\ln \left( \frac{2\sqrt{2}}{r_s} \right) 
\e
-
\frac{1}{16}\frac{\e|\e|}{\Ef}
+
\frac{5 }{48\sqrt{2}\pi }\frac{ \e^3}{r_s\Ef^2} \ln \left( \frac{ r_s \Ef}{ |\e| }\right)
+
\frac{-41+75\ln 2}{288 \sqrt{2} \pi }\frac{ \e^3}{r_s\Ef^2}.
\end{aligned}
\end{align}

\subsection{3D electron self-energy}\label{sec:3Dresult}

 Starting from the general formulas Eq.~\ref{eq:Sig6}, one can derive the 3D electron self-energy in a way analogous to the one presented in Sec.~\ref{sec:2Dresult} in the case of 2D. In this section, without giving the details of the calculation, we present directly the explicit expressions for the 3D electron self-energy on the mass shell, which are valid to the leading order in $r_s$, and to several orders in $\e/\Ef$ and $T/\Ef$. The 3D calculations follow the general procedure given above in depth for the corresponding 2D theory. To simplify the calculation, we consider a range of temperature (energy) slightly different from the one in the 2D case:  $r_s\ll {T}/{\Tf} \ll \sqrt{r_s} $ ($r_s\ll {\e}/{\Ef} \ll \sqrt{r_s} $).
 
 In 3D, for arbitrary energy-to-temperature ratio, the imaginary part of the on-shell self-energy acquires the form of 
\begin{align}\label{eq:3DimSig}
\begin{aligned}
	\im \Sigma^{(R)}(\e,T)=\,&	
	-
	\frac{1}{32} (\frac{\pi^4}{12})^{1/3} 
	\sqrt{r_s}
	\frac{1}{\Ef}
	\left( \pi^2T^2+\e^2\right) 
	-
	\frac{\pi}{8} C_0 \frac{T^3}{\Ef^2} 
	\left[ \zeta(3)-\frac{1}{2} \left( \Li_3(-e^{-\frac{\e}{T}})+  \Li_3(-e^{ \frac{\e}{T}}) \right)   \right] , 
\end{aligned}
\end{align}
where $C_0$ is a constant defined as
\begin{align}
	C_0
	=
	\int_{0}^{1} dy
	\frac{1}{y^2}
	\left[ 
	\dfrac{	1}
	{\left( 1
		-\frac{y}{2}
		\ln 
		\left(  \dfrac{1+y}{1-y} \right)  \right)^2+ \left( \frac{\pi}{2} y\right)^2 }
	-1
	\right] 
	-1
	\approx
	-1.5326.
\end{align}
From the equation above, one can find $\im \Sigma^{(R)}$ in both the low-energy limit and the low-temperature limit:
\begin{subequations}\label{eq:3D-ImSig}
\begin{align}
	&\begin{aligned}\label{eq:3D-ImSig-<}
	\im \Sigma^{(R)}(|\e | \ll T)=\,&	
	-
	\frac{1}{32} (\frac{\pi^4}{12})^{1/3} 
	\pi^2  
	\sqrt{r_s}
	\frac{T^2}{\Ef}
	-
	\frac{7\pi}{32} \zeta(3) C_0  \frac{T^3}{\Ef^2},   
	\end{aligned}
	\\
	&\begin{aligned}\label{eq:3D-ImSig->}
	\im \Sigma^{(R)}(|\e| \gg T)=\,&	
	-
	\frac{1}{32} (\frac{\pi^4}{12})^{1/3} 
	\sqrt{r_s}
	\frac{\e^2}{\Ef}
	-
	\frac{\pi}{96}  C_0\frac{|\e|^3}{\Ef^2}. 
	\end{aligned}
\end{align}
\end{subequations}

Similarly, we evaluate  the real part of the 3D on-shell self-energy in the case where $\e$ and $T$ are arbitrary with respect to each other, and find
\begin{align}\label{eq:3DreSig}
\begin{aligned}
	\re \Sigma^{(R)} (\varepsilon,T)
	=\,&
	\frac{1}{3\pi^{4/3}} (\frac{2}{3})^{2/3} r_s
	\ln \left( \frac{3\pi^2}{2r_s^{3/2}}\right) 
	\e
		-
	\frac{4-\pi^2}{192}	\ln 
	\left( \frac{1}{2(\frac{16}{3\pi^2})^{1/3} \sqrt{r_s}} \frac{T}{\Ef}\right) 	
	\frac{T^3}{ \Ef^2} 
	\left( 
	\frac{\e^3}{T^3}+\pi^2\frac{\e}{T} 
	\right) 
	\\
	&
	-
	\left[ 
	\frac{4-\pi^2}{384}
	\left( 3-2\gamma_E-2\ln 2 \right) 
	-
	\frac{2}{3} C_1
	\right] 
	\frac{T^3}{ \Ef^2}
	\left( \frac{\e^3}{T^3}+\pi^2\frac{\e}{T} \right)
	\\
	&
	-
	\frac{4-\pi^2}{32}	\frac{T^3}{ \Ef^2}
	\left[ \partial_s \Li_s (-e^{-\frac{\e}{T}}) - \partial_s \Li_s (-e^{\frac{\e}{T}}) \right] \bigg\vert_{s=3},
\end{aligned}
\end{align}
where
\begin{align}
\begin{aligned}
	C_1
	\equiv	
	\frac{1}{32}
	\int_{0}^{1} dy
		\frac{1}{y^3}
	\left[ 
	\dfrac{1
		-\frac{y}{2}
		\ln 
		\left(  \dfrac{1+y}{1-y} \right)  }
	{\left( 1
		-\frac{y}{2}
		\ln 
		\left( \dfrac{1+y}{1-y} \right) \right)^2+ \left( \frac{\pi}{2} y\right)^2 }
	-
	\left( 1+\frac{4-\pi^2}{4} y^2\right) 
	\right] 
	+\frac{-16+3\pi^2-4(4-\pi^2) \ln 2 }{512}
	\approx 0.059.
\end{aligned}
\end{align}
In the limits where $|\e|/T\ll 1$ and $|\e|/T \gg 1$, Eq.~\ref{eq:3DreSig} reduces to, respectively,
\begin{subequations}
\begin{align}
	&\begin{aligned}\label{eq:3D-ReSig-<}
	\re \Sigma^{(R)} (T \gg |\varepsilon|  )
	=\,&
	\frac{1}{3\pi^{4/3}} (\frac{2}{3})^{2/3} r_s
	\ln \left( \frac{3\pi^2}{2r_s^{3/2}}\right) 
	\e
	-
	\frac{4-\pi^2}{192}	\pi^2 
	\ln \left( \frac{1}{2(\frac{16}{3\pi^2})^{1/3} \sqrt{r_s}} \frac{T}{\Ef}\right) 	
	\frac{T^2 \e}{ \Ef^2}
	\\
	&
	-
	\left[ 
	\frac{4-\pi^2}{384}
	\left( 3-2\gamma_E-2\ln 2 \right) 
	-
	\frac{2}{3} C_1
	+
	\frac{4-\pi^2}{32\pi^2}	
	\left( \frac{\pi^2}{6} \ln 2+\zeta'(2)\right) 
	\right] 
	\pi^2
	\frac{T^2 \e }{ \Ef^2},
	\end{aligned}
	\\
	&\begin{aligned}\label{eq:3D-ReSig->}
	\re \Sigma^{(R)} (|\varepsilon |\gg T)
	=\,&
	\frac{1}{3\pi^{4/3}} (\frac{2}{3})^{2/3} r_s
	\ln \left( \frac{3\pi^2}{2r_s^{3/2}}\right) 
	\e
		-
	\frac{4-\pi^2}{192}	\ln \left( \frac{1}{2(\frac{16}{3\pi^2})^{1/3} \sqrt{r_s}}  \frac{|\e|}{\Ef}\right) 	
	\frac{\e^3}{ \Ef^2} 
	\\
	&
	+
	\left[ 
	\frac{4-\pi^2}{576}
	\left( 1+3\ln 2 \right) 
	+
	\frac{2}{3} C_1
	\right] 
	\frac{\e^3}{ \Ef^2}.
	\\
	\end{aligned}
\end{align}
\end{subequations}

\subsection{Discussion of the analytical results}

It may be worthwhile to summarize our analytical findings for the self-energy at low temperatures and energies (and to the leading-order in $r_s$).

Both in 2D and 3D, $\im \Sigma^{(R)}$ goes as $T^2$ for $\varepsilon =0$ and as $\varepsilon^2$ for $T=0$ (with an additional log correction in 2D) in the leading order, as is already well-known.  This establishes the perturbative stability of the Fermi surface, implying that both 3D and 2D interacting systems are Fermi liquids, in contrast to interacting 1D fermions.  The subleading terms in the 3D $\im \Sigma^{(R)}$ are $O(T^3)$ for $\varepsilon=0$ and $O(\varepsilon^3)$ for $T=0$.  In 2D, however, the corresponding subleading terms are $O(T^2)$ and $O(\varepsilon^2)$, respectively since the leading order terms go as $O(T^2\ln T)$ and $O(\varepsilon^2 \ln  \varepsilon)$.  The next order terms in 2D are cubic, as expected.

When energy and temperature are comparable, the results for $\im \Sigma^{(R)}$ are complicated, involving logarithmic integrals in $\exp (-\varepsilon/T)$ in addition to powers of $\varepsilon$ and $T$ along with log factors in 2D.  In 3D (Eq.~\ref{eq:3DimSig}),  $\im \Sigma^{(R)}$ goes as $(\varepsilon ^2 + \pi^2 T^2)$ plus term involving logarithmic integrals.  In 2D (Eq. 2.69), $\im \Sigma^{(R)}$ for small, but comparable,  $\varepsilon$ and $T$, goes as $(\varepsilon^2 + \pi^2 T^2) \ln (T/\Ef)$ with additional terms
involving $O(T^2)$, $O(\varepsilon ^2)$, and logarithmic integrals.

The analytical behavior of $\re \Sigma^{(R)}$ is as follows.

In 3D, for $\varepsilon\ll T$, $\re \Sigma^{(R)}$ goes as $O(\varepsilon) + O (\varepsilon  T^2 \ln T) + O(\varepsilon T^2)$, whereas for $\varepsilon\gg T$, it goes as $O(\varepsilon) + O(\varepsilon ^3 \ln \varepsilon) + O (\varepsilon ^3)$.  In 2D, for $\varepsilon\ll T$, $\re \Sigma^{(R)}$ goes as $O(\varepsilon) + O(\varepsilon T) + O (\varepsilon  T^2 \ln T) + O (\varepsilon  T^2)$, whereas for $\varepsilon\gg T$, it goes as $O(\varepsilon) + O(\varepsilon^2) + O(\varepsilon^3 \ln  \varepsilon) + O (\varepsilon^3)$.

When energy and temperature are comparable, but both small, the behavior of $\re \Sigma^{(R)}$ is complicated with the appearance of logarithmic integrals similar to the situation for $\im \Sigma^{(R)}$ discussed above.  In 3D, $\re \Sigma^{(R)}$ then goes as $O(\varepsilon) + O ( [\varepsilon ^3 + \pi^2 \varepsilon T^2] \ln T) + O (\varepsilon^3 + \pi^2 \varepsilon T^2) $ plus terms involving logarithmic integrals as in Eq.~\ref{eq:3DreSig}.  In 2D, for general small values of energy and temperature, $\re \Sigma^{(R)}$ behaves as $O(\varepsilon) + O ([\pi^2 T^2\varepsilon + \varepsilon ^3] \ln T) + O (\pi^2 T^2\varepsilon + \varepsilon ^3)$ plus several terms involving logarithmic integrals as shown in Eq.~\ref{eq:ReSig}.

The presence of various logarithmic terms and combinations of powers of $T$ and $\varepsilon$ along with logarithmic integrals made the calculation of the self-energy a challenge for arbitrary (but small) energy and temperature even in the $r_s\ll 1$ limit for the last 60 years~\cite{AGD}, which we finally managed to resolve~\cite{PRB}.

The $O(\varepsilon^2)$ or $O(\varepsilon^2 \ln \varepsilon)$ asymptotic behavior of $\im \Sigma^{(R)}$ in 3D or 2D respectively assures the existence of a Fermi surface at $\varepsilon=0$ since $\re \Sigma^{(R)}$ always goes as $O(\varepsilon)$.  The question we address in the rest of this paper is what happens at finite temperature and energy where $\im \Sigma^{(R)}$ in principle could be larger than energy or temperature itself, indicating that the quasiparticle picture fails there.  

Our goal is to ascertain the domain of validity of the Fermi liquid theory and the associated quasiparticle picture by comparing the quasiparticle energy $\varepsilon$ with the quasiparticle damping defined by the magnitude of $\im \Sigma^{(R)} (\varepsilon)$, both in 3D and 2D, contrasting the two cases.

\section{Results}

\subsection{Applicability of the FL theory}\label{sec:applicability}

The Fermi liquid theory describes interacting Fermi systems at low temperatures and excitation energies in terms of quasiparticles - long-lived elementary excitations which are adiabatically connected to the excitations of noninteracting systems.
Therefore it relies on the existence of well-defined quasiparticles, whose damping rates, determined by the imaginary part of the self-energy, should be small compared with their energies.
 More specifically, the quasiparticle description is valid when the magnitude of the imaginary part of the self-energy $|\im \Sigma^{(R)}(\e)|$ is small compared with $\e$.
 To the leading order, $\im \Sigma^{(R)}(\e)$ scales as $\e^2\ln \e$ and $\e^2$, respectively, in 2D and 3D (see Refs.~\cite{Galitskii1958,Quinn1958,Chaplik,Quinn1982,Zheng,Li2013}, as well as Eqs.~\ref{eq:ImSig->} and~\ref{eq:3D-ImSig->}). Therefore,  at sufficiently low energy $\e$, the criterion $|\im \Sigma^{(R)}(\e)|<\e$ is satisfied and the quasiparticle is well defined.
 However, at higher energy or temperature, this might no longer be the case. 
In this section, we use the previously obtained analytical expressions for the electron self-energy to determine the regime in which the quasiparticle description is applicable (invalid) - called Fermi-liquid (FL) or non-Fermi liquid (NFL) regime in the following.
 We compute the on-shell $\im \Sigma^{(R)}(\e)/\e$ at zero temperature as well as $\im \Sigma^{(R)}(T)/T$ at the Fermi level (i.e. $\varepsilon=0$), in an attempt to determine the crossover energy $\e_c$ and crossover temperature $T_c$ below which the quasiparticles are well defined.
$\e_c$ and $T_c$ separate the FL and NFL regimes, and are determined by the conditions:
\begin{align}\label{eq:TEC}
\begin{aligned}
-\im \Sigma^{(R)}(\e_c,T=0)/\e_c=1,
\qquad
-\im \Sigma^{(R)}(\e=0,T_c)/T_c=1.
\end{aligned}
\end{align}
Of course, the precise magnitudes of $\varepsilon_c$ and $T_c$ will depend on the expressions for the self-energy we use, and the leading-order and sub-leading-order theories may give different results, but our interest is in understanding the qualitative trends being mindful of the fact that the analytical theory is meaningful only up to energy/temperature where the subleading terms are smaller than the leading terms and the constraint $\varepsilon, T < \Ef$ applies.
Strictly speaking, the analytical expressions for the electron self-energy presented in the previous section are derived within RPA, and are valid in the high-density, low-energy and low-temperature regime.
However, it has been found that the RPA approximation works reasonably well even outside the high density regime.
For example, in Ref.~\cite{Rice}, it has been shown that, for an electron gas at metallic densities,  the effective mass, Pauli spin susceptibility and compressibility obtained from RPA are in good agreement with the experiment.
We note that $\e/\Ef r_s$ and $T/\Ef r_s$ ($\e/\Ef \sqrt{r_s}$ and $T/\Ef \sqrt{r_s}$) are used, apart from $r_s$, as the small expansion parameters in the calculation of the 2D (3D) self-energy. Larger $r_s$ therefore means wider energy and temperature ranges of applicability (for not too large value of $r_s$).
For these reasons, in this section, we assume  that the previously obtained analytical expressions for the self-energy apply to arbitrary $r_s$, $\e/\Ef$ and $T/\Ef$, and use them to estimate when the quasiparticle description breaks down.
Our goal is to determine the energy/temperature regime where the Fermi liquid theory applies within our approximations.

\begin{figure}[t!]
	\centering
	\includegraphics[width=0.7\linewidth]{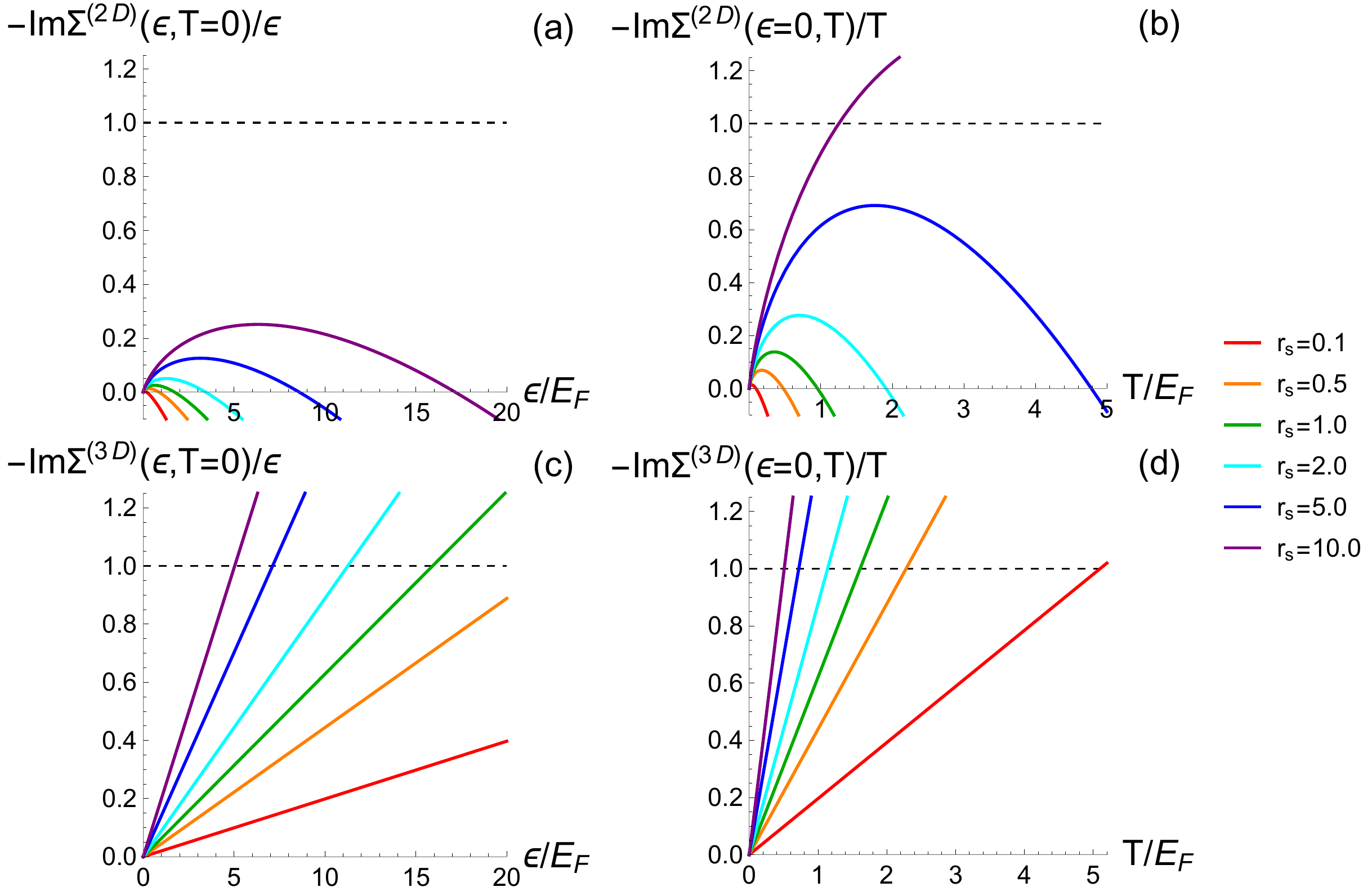}
	\caption{(a) $-\im \Sigma^{(R)}(\e,T)/\e$ as a function of $\e/\Ef$ at $T=0$ in 2D; (b) $-\im \Sigma^{(R)}(\e,T)/T$ as a function of $T/\Ef$ at $\e=0$ in 2D; (c) and (d) are same as (a) and (b), respectively, but in $d=3$ dimensions. 
	(a) [(c)] is obtained using the zero temperature self-energy expression Eq.~\ref{eq:ImSig->} (Eq.~\ref{eq:3D-ImSig->}) up to the $\e^2$ term, while (b) [(d)] is from zero energy self-energy expression Eq.~\ref{eq:ImSig-<} (Eq.~\ref{eq:3D-ImSig-<}) up to the $T^2$ term.
	In all four panels, we consider $r_s=0.1, 0.5, 1.0, 2.0, 5.0, 10.0$ represented by solid lines with different colors.
	The horizontal dashed lines take the value of $1$ and separate the FL and NFL regimes where the quasiparticle description is applicable and invalid, respectively.
	}
	\label{fig:p-2}
\end{figure}

In Fig.~\ref{fig:p-2}(a) (Fig.~\ref{fig:p-2}(c)), the ratio $-\im \Sigma^{(R)}(\e, T)/\e$ at $T=0$ is plotted as a function of the dimensionless energy $\e/\Ef$ for 2D (3D) electron systems for various interaction parameter $r_s$ values. 
For Fig.~\ref{fig:p-2}(a), we apply Eq.~\ref{eq:ImSig->}, the analytical expression for zero temperature $\im \Sigma^{(R)}$ in 2D, and retain the leading order $\e^2\ln \e$ term as well as the subleading $\e^2$ term, while for Fig.~\ref{fig:p-2}(c), only the leading order $\e^2$ term in the zero temperature expression for the imaginary part of the 3D self-energy  (Eq.~\ref{eq:3D-ImSig->}) is used.
Figs.~\ref{fig:p-2}(b) and~\ref{fig:p-2}(d) are plotted in a similar way. Instead of  $-\im \Sigma^{(R)}(\e, T=0)/\e$, we plot $-\im \Sigma^{(R)}(\e=0,T)/T$ as a function of $T/\Ef$ for the same set of $r_s$ values for both 2D (Fig.~\ref{fig:p-2}(b)) and 3D (Fig.~\ref{fig:p-2}(d)) systems. For these two figures, we apply Eq.~\ref{eq:ImSig-<} and Eq.~\ref{eq:3D-ImSig-<}, which give the zero-energy form of $\im \Sigma^{(R)}$ for $d=2$ and $d=3$, respectively, and neglect the highest order $T^3$ terms in both equations.
 

In Figs.~\ref{fig:p-2}(a) and~\ref{fig:p-2}(b), which depict the 2D case, almost all curves lie below the horizontal dashed line with the value of $1$ (except for the one corresponds to $r_s=10$ in Fig.~\ref{fig:p-2}(b)). This means that the conditions $-\im \Sigma^{(R)}(\e, T=0)/\e<1$ and $-\im \Sigma^{(R)}(\e=0, T)/T<1$ always hold in the observed regime. 
However, as $\e/\Ef$ $(T/\Ef)$ increases, the self-energy expression used to plot Fig.~\ref{fig:p-2}(a) (Fig.~\ref{fig:p-2}(b)) is no longer valid, and $-\im \Sigma^{(R)}(\e, T=0)/\e$ $\left( -\im \Sigma^{(R)}(\e=0, T)/T\right) $ becomes negative for large enough $\varepsilon$ ($T$) for almost all $r_s$ considered.
This implies that the corresponding approximation breaks down at high energy, but quasiparticles remain well-defined up to the energy cut off where the theory is valid.
In Fig.~\ref{fig:p-2}(b), the  purple curve which represents the case of $r_s=10$ intersects with $-\im \Sigma^{(R)}/T=1$ line (dashed line) at $T_c(r_s=10)\approx1.25\Ef$, above which the quasiparticle is no longer well defined (for this value of $r_s$).
If we smoothly extrapolate the 2D results in Figs.~\ref{fig:p-2}(a) and~~\ref{fig:p-2}(b) from their low energy  monotonic behavior (before the pathological maxima induced by the failure of the expansion), then the results for different $r_s$ values all smoothly cross the dashed line (i.e. unity) for increasing $\varepsilon/\Ef$ and $T/\Tf$ with decreasing $r_s$, implying that the Fermi liquid regime is larger for lower $r_s$, which is understandable since lower $r_s$ indicates weaker interactions.  

In 3D (Figs.~\ref{fig:p-2}(c) and~\ref{fig:p-2}(d)), $-\im \Sigma^{(R)}(\e, T=0)/\e$ ($-\im \Sigma^{(R)}(\e=0, T)/T$) from the leading order zero temperature (zero energy) self-energy expression remains positive for all $\e$ ($T$). Furthermore, for each $r_s$, there exists a $\e_c$ ($T_c$) above which the ratio $-\im \Sigma^{(R)}(\e, T=0)/\e$ $\left( -\im \Sigma^{(R)}(\e=0, T)/T\right) $ exceeds $1$,  signaling the break down of the quasiparticle description. 
Substituting the leading order terms in Eq.~\ref{eq:3D-ImSig->} and \ref{eq:3D-ImSig-<} into Eq.~\ref{eq:TEC},  we have
\begin{align}\label{eq:TEC-3D}
	\begin{aligned}
		\frac{\e_c^{(3D)}}{\Ef}=\,
		32 (\frac{12}{\pi^4})^{1/3} 
		\frac{1}{\sqrt{r_s}},
		\qquad
		\frac{T_c^{(3D)}}{\Ef}=\,
		32 (\frac{12}{\pi^{10}})^{1/3} 
		\frac{1}{\sqrt{r_s}}.
	\end{aligned}
\end{align}
From this result, one can see that $\e_c^{(3D)}/\Ef$ and $T_c^{(3D)}/\Ef$ have a $r_s$-dependence of $1/\sqrt{r_s}$ and decay with increasing $r_s$,  in accordance with Figs.~\ref{fig:p-2}(c) and~\ref{fig:p-2}(d).
The regime of Fermi liquid validity shrinks in energy with increasing $r_s$ consistent with increasing interaction strength in the system.
We emphasize that this result is only an approximation from the leading order 3D self-energy expressions valid for a small range of temperatures and energies.

\begin{figure}[t!]
	\centering
	\includegraphics[width=0.7\linewidth]{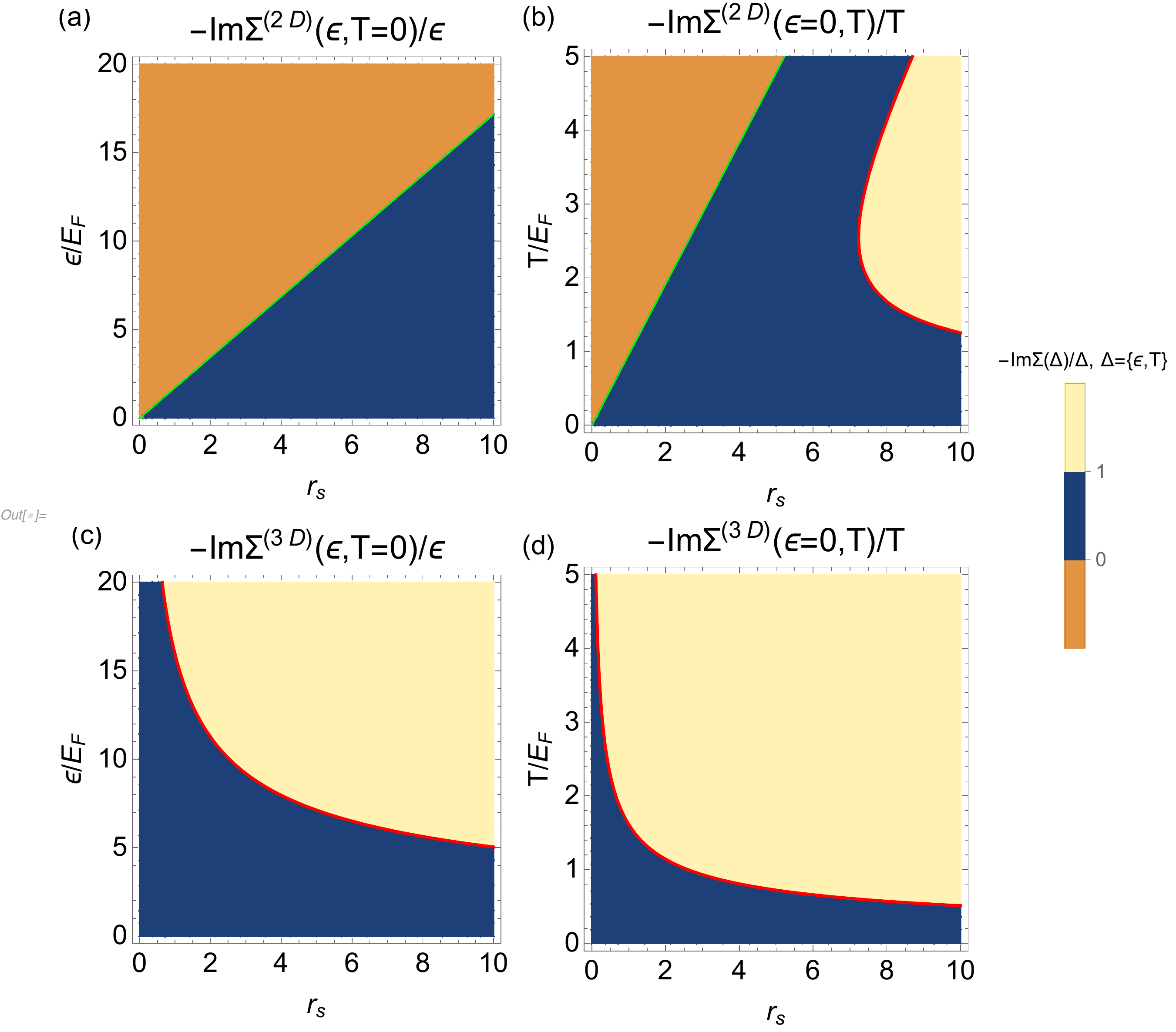}
	\caption{Plots of $-\im \Sigma^{(R)}(\e, T=0) /\e$ in the $(r_s, \e/\Ef)$-plane (left panels), and $-\im \Sigma^{(R)} (\e=0,T)/T$  in the $(r_s, T/\Ef)$-plane (right panels) in 2D (upper panels) and 3D (lower panels). 
		For panels (a)-(d),
		Eqs.~\ref{eq:ImSig->},~\ref{eq:ImSig-<},~\ref{eq:3D-ImSig->} and~\ref{eq:3D-ImSig-<} are used, in respective order, and the highest order $T^3$ or $\e^3$ terms in these equations are neglected.
		In the left (right) panels, the FL regime where $0<-\im \Sigma(\e, T=0) /\e <1$ ($0<-\im \Sigma^{(R)}(\e=0, T) /T <1$) is indicated by the blue area, whereas the NFL regime where $-\im \Sigma^{(R)}(\e, T=0) /\e >1$ ($-\im \Sigma^{(R)}(\e=0, T) /T >1$) is represented by the yellow area. Their boundary is indicated by red contour line which corresponds to $\e_c$ ($T_c$) defined in Eq.~\ref{eq:TEC}.
		In the region colored in orange, which is separated from the rest by green contour with the value of zero, the approximated self-energy expressions are no longer valid and lead to a positive value of $\im \Sigma^{(R)}$.}
	\label{fig:cp-2}
\end{figure}

Using the same formulas for the imaginary part of the self-energy (given by an expansion to order $\e^2$ for the zero temperature case and to order $T^2$ for the zero energy case), in  Fig~\ref{fig:cp-2}, we show contour plots of $\im \Sigma^{(R)}/\e$ at $T=0$ in the $(r_s,\e/\Ef)$-plane (left panels), as well as  $-\im \Sigma^{(R)}/T$ at $\e=0$ in the $(r_s,T/\Ef)$-plane (right panels) for $d=2$ (upper panels) and $d=3$ (lower panels). In the left (right) panels, the blue areas are used to indicate the FL regimes where $0<-\im \Sigma^{(R)} /\e <1$ ($0<-\im \Sigma^{(R)} /T<1$), whereas the yellow areas correspond to the NFL regimes where $-\im \Sigma^{(R)} /\e >1$ ($-\im \Sigma^{(R)} /T>1$). 
They are separated by the red contour lines which take the value of one and correspond to $\e_c$ ($T_c$) defined by Eq.~\ref{eq:TEC}. As shown in Figs.~\ref{fig:cp-2}(c) and~\ref{fig:cp-2}(d), $\e_c^{(3D)}/\Ef$ and $T_c^{(3D)}/\Ef$  decreases monotonically with increasing $r_s$ in 3D, consistent with Eq.~\ref{eq:TEC-3D}.
In Figs.~\ref{fig:cp-2}(a) and~\ref{fig:cp-2}(b), there exist regimes where the self-energy expressions are no longer valid and the resulting 2D $\im \Sigma^{(R)}$ is positive. We use the orange areas to indicate such regimes, where the current self-energy expressions are unable to tell if the quasiparticle is well defined or not.
Furthermore, using the current approximation for $\im \Sigma^{(R)}$,  the solution to Eq.~\ref{eq:TEC} - $\e_c^{(2D)}$ (or $T_c^{(2D)}$ for small $r_s$) does not exist (see Figs.~\ref{fig:cp-2}(a) and~\ref{fig:cp-2}(b)) . In this case, $-\im \Sigma^{(R)} /\e<1$ ($-\im \Sigma^{(R)} /T<1$) always holds in the region where this formula is applicable.
We mention that if we use extrapolations of the 2D self-energy from their low energy behavior so that the pathological behavior of the imaginary self-energy decreasing at higher energies and temperatures is eliminated (so that the orange region in the figures disappears), then the 2D results in Fig.~\ref{fig:cp-2} look qualitatively the same as the 3D results with the orange regions in Fig.~\ref{fig:cp-2}(a) and~\ref{fig:cp-2}(b) mostly becoming blue.

\begin{figure}[t!]
	\centering
	\includegraphics[width=0.5\linewidth]{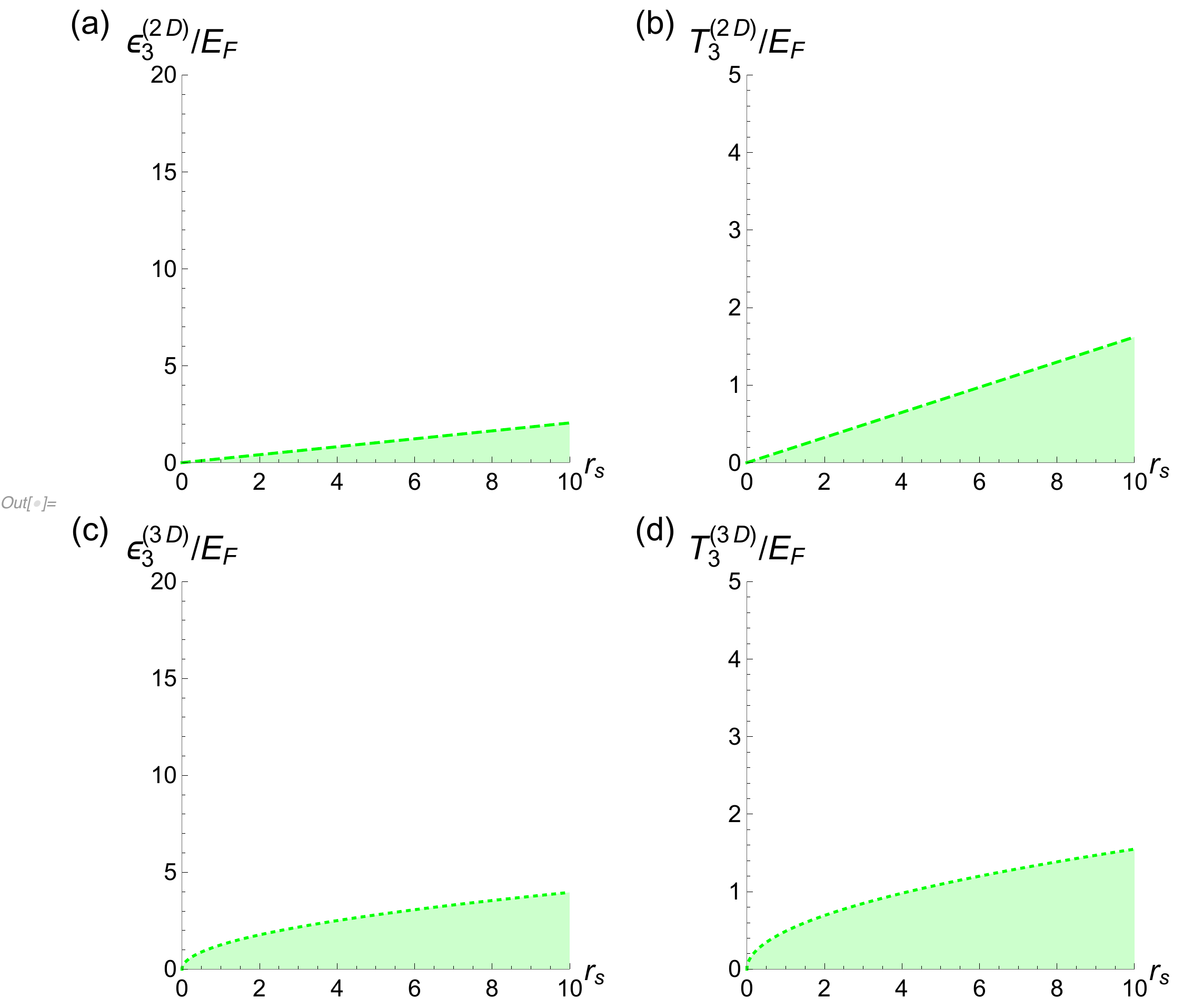}
	\caption{
		The regimes (green areas) where the applicable criterion - $E^3(T^3)$ term must be smaller than $E^2 (T^2)$ term - is satisfied for self-energy expressions Eq.~\ref{eq:ImSig->}  (panel a), Eq.~\ref{eq:ImSig-<}  (panel b), Eq.~\ref{eq:3D-ImSig->} (panel c) and Eq.~\ref{eq:3D-ImSig-<} (panel d).}
	\label{fig:v}
\end{figure}

As explained earlier, the self-energy expressions used to obtain Figs.~\ref{fig:p-2} and~\ref{fig:cp-2} are applicable to a small range of energies and temperatures because of the leading order nature of the analytical expansion.
We now retain the $T^3$ and $\e^3$ terms in the self-energy formulas Eqs.~\ref{eq:ImSig->},~\ref{eq:ImSig-<},~\ref{eq:3D-ImSig->} and~\ref{eq:3D-ImSig-<} to evaluate $\im \Sigma^{(R)}/\e$ and $\im \Sigma^{(R)}/T$ and to analyze the applicable regime of the FL quasiparticle description. 
We note that these expressions have a wider range of applicability compared with the one used in Figs.~\ref{fig:p-2} and~\ref{fig:cp-2}, but are only valid when the $T^3$ ($\e^3$) term is smaller than the $T^2$ ($\e^2$) term.  
We find that the leading upper bounds for the regimes satisfying this condition for the 2D and 3D cases are given by, respectively,
\begin{align}\label{eq:et3}
\begin{aligned}
	\frac{\e_3^{(2D)}}{\Ef}
	=\,&
	\frac{3\sqrt{2} \left( \ln 4 -1\right) }{8 } r_s,
	\qquad
	&
	\frac{T_3^{(2D)}}{\Ef}
	=\,&
	\frac{\pi^2\left(6 + \ln 2\pi^3 -36 \ln \mathrm{A} \right) }{42\sqrt{2}\zeta(3) }r_s,
	\\
	\frac{\e_3^{(3D)}}{\Ef}
	=\,&
	\frac{ (\frac{9\pi}{4})^{1/3} }{ C_0} \sqrt{r_s},
	\qquad
	&
	\frac{T_3^{(3D)}}{\Ef}
	=\,&
	\frac{ (\frac{\pi^7}{12})^{1/3}}{7\zeta(3) C_0} \sqrt{r_s}.
\end{aligned}
\end{align}
We note that, in 2D, within the applicability range $\e<\e_3^{(2D)}$ ($T<T_3^{(2D)}$), the leading order $\e^2\ln\e$ ($T^2\ln T$) term is also larger than $\e^2$ ($T^2$) term.
$\e_3$ (left panels) and $T_3$ (right panels) are plotted by the dashed green curves in Fig.~\ref{fig:v} for 2D (upper panels) and 3D (lower panels) systems. The green areas in this figure represent the regimes where $E^3 (T^3)$ term in the self-energy expressions is always smaller than $E^2 (T^2)$ term.

\begin{figure}[t!]
	\centering
	\includegraphics[width=0.7\linewidth]{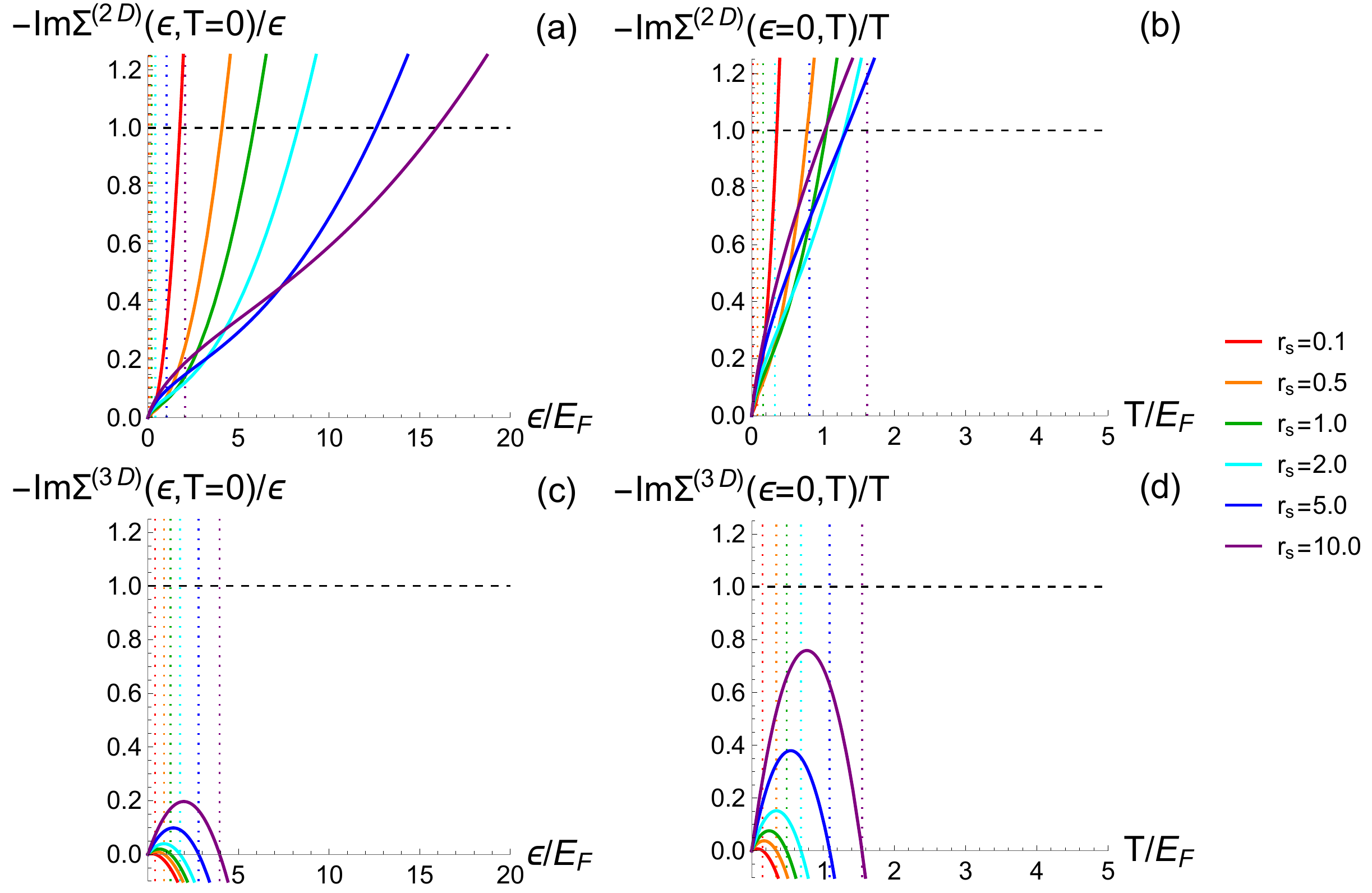}
	\caption{The same as Fig.~\ref{fig:p-2} but retaining the highest order $T^3$ and $\e^3$ terms in Eqs.~\ref{eq:ImSig->},~\ref{eq:ImSig-<},~\ref{eq:3D-ImSig->} and~\ref{eq:3D-ImSig-<}.}
	\label{fig:p-3}
\end{figure}

\begin{figure}[t!]
	\centering
	\includegraphics[width=0.7\linewidth]{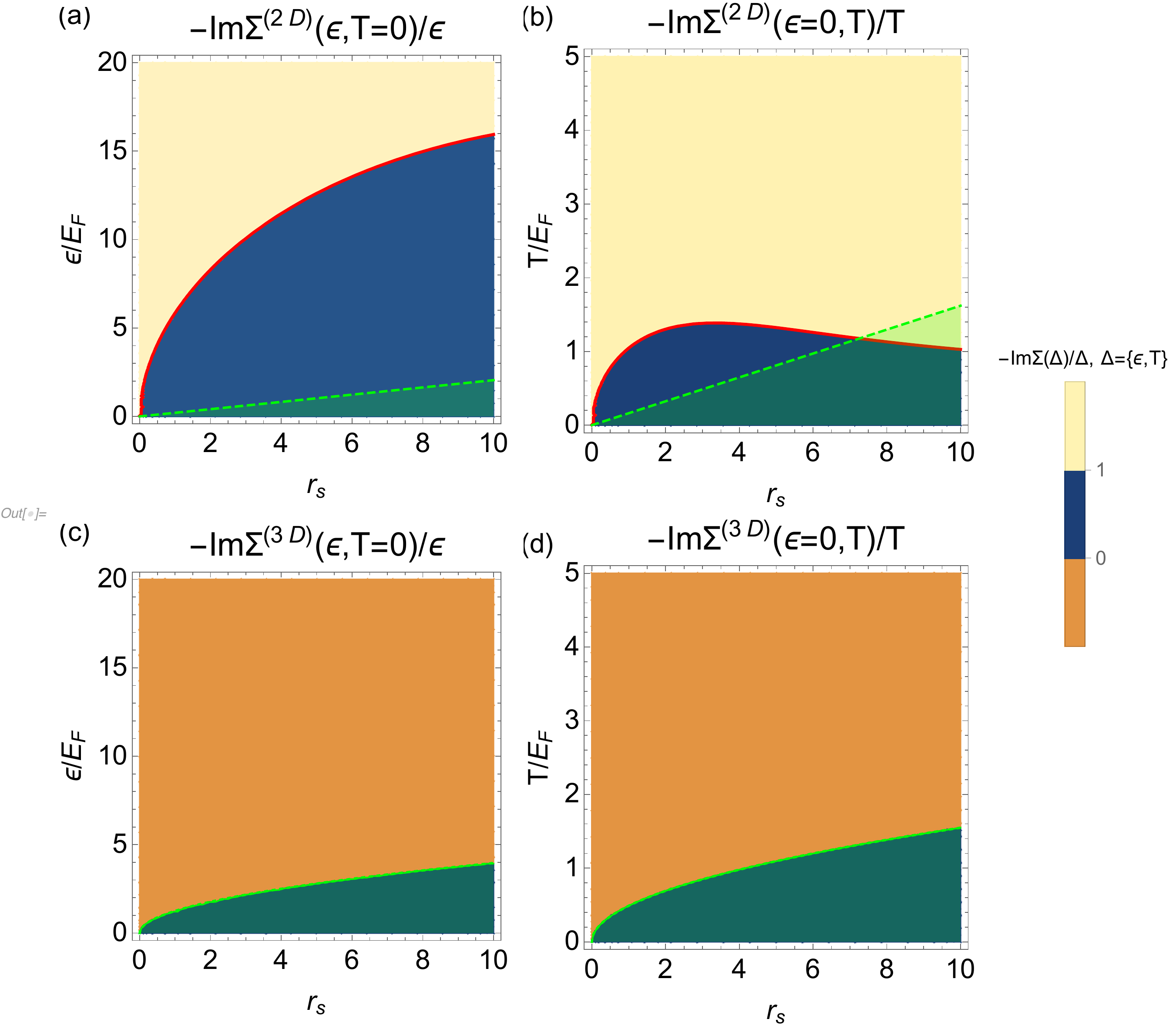}
	\caption{The same as Fig.~\ref{fig:cp-2} but using the self-energy expressions which are valid up to $\e^3$ and $T^3$ (i.e. keeping all terms in Eqs.~\ref{eq:ImSig->},~\ref{eq:ImSig-<},~\ref{eq:3D-ImSig->} and~\ref{eq:3D-ImSig-<}).}
	\label{fig:cp-3}
\end{figure}

In Figs.~\ref{fig:p-3} and~\ref{fig:cp-3}, we replot $-\im \Sigma^{(R)}(\e,T=0)/\e$ and $-\im \Sigma^{(R)}(\e=0,T)/T$ as in Figs.~\ref{fig:p-2} and~\ref{fig:cp-2}, but with the self-energy expressions which include the $T^3 (\e^3)$ term.
As in Fig.~\ref{fig:p-2}, different curves in Fig.~\ref{fig:cp-3} correspond to different values of $r_s$. For each curve, we use a vertical dotted line of the same color to indicate the upper limit $\e_3$ or $T_3$ (Eq.~\ref{eq:et3}) of the applicable range for the corresponding self-energy expression.
We notice that, for the 2D case, these curves exhibit nonmonotonic dependence on $r_s$, which will disappear if we incorporate the $r_s$ dependence of the Fermi energy ($E\sim r_s^{-2}$) and plot in fixed units instead (see Figs.~\ref{fig:F}(a) and~\ref{fig:F}(b)).
In the observed regime, for the 3D case, the ratios $-\im \Sigma^{(R)}(\e,T=0)/\e$ and $-\im \Sigma^{(R)}(\e=0,T)/T$ are always smaller than one and become negative (similar to what happens for 2D results in Fig.~\ref{fig:p-2}) outside the applicable ranges of the self-energy expressions (see Figs.~\ref{fig:p-3}(c) and~\ref{fig:p-3}(d)), whereas for the 2D case these ratios exceed one at higher energies and temperatures (see Figs.~\ref{fig:p-3}(a) and~\ref{fig:p-3}(b)).

In Fig.~\ref{fig:cp-3}, the blue, yellow and orange areas correspond to, respectively, the regimes where the ratio $-\im \Sigma^{(R)}(\e,T=0)/\e$ ($-\im \Sigma^{(R)}(\e=0,T)/T$) lies within $(0,1)$,  stays above one, and remains negative (i.e. theoretically inapplicable),  as in Fig.~\ref{fig:cp-2}.
We also add a green area in each panel to indicate the region where the corresponding self-energy expression is applicable (see Eq.~\ref{eq:et3}).
In Figs.~\ref{fig:cp-3}(a) and~\ref{fig:cp-3}(b) which are associated with the 2D case, we find both FL and NFL regimes,  indicated by the blue and yellow areas, respectively. The red contour lines take the value of one and represent $\e_c^{(2D)}$ (Fig.~\ref{fig:cp-3} (a)) or $T_c^{(2D)}$ (Fig.~\ref{fig:cp-3} (b)). 
As one can see  from these figures, over some range of $r_s$ values, $\e_c^{(2D)}/\Ef$ ($T_c^{(2D)}/\Ef$) grows with increasing $r_s$ which seems contrary to a naive expectation that the quasiparticle description breaks down at lower energy and temperature for larger interaction strength. However, taking into account the $r_s$ dependence of $\Ef $, we find that $\e_c^{(2D)}$ and $T_c^{(2D)}$ in fixed units decay with increasing $r_s$ as expected. See Figs.~\ref{fig:F}(c) and~\ref{fig:F}(d) where we replot Figs.~\ref{fig:cp-3}(a) and~\ref{fig:cp-3}(b) in fixed units.
We also note that $\e_c^{(2D)}$ and $T_c^{(2D)}$ stay outside the applicable regime of the self-energy expression (green area), and are therefore not reliable.
This result shows that within the applicable regime, the FL quasiparticle description is valid.
In the lower panels of Fig.~\ref{fig:cp-3} which depict the 3D cases, the yellow areas disappear. Instead, one finds orange areas which correspond to the regimes where the associated self-energy expressions are invalid and yield $\im \Sigma^{(R)}>0$. 
As before, we have no information about whether the quasiparticle is well defined or not in the orange areas.
The current analytical theory is simply inapplicable in the orange areas.
It is, in principle, possible that the Fermi liquid theory remains valid with well-defined quasiparticles for all energies and temperatures since our analytical theory cannot access arbitrary quasiparticle energy and temperature well above $\Ef$.

\begin{figure}[t!]
	\centering
	\includegraphics[width=0.7\linewidth]{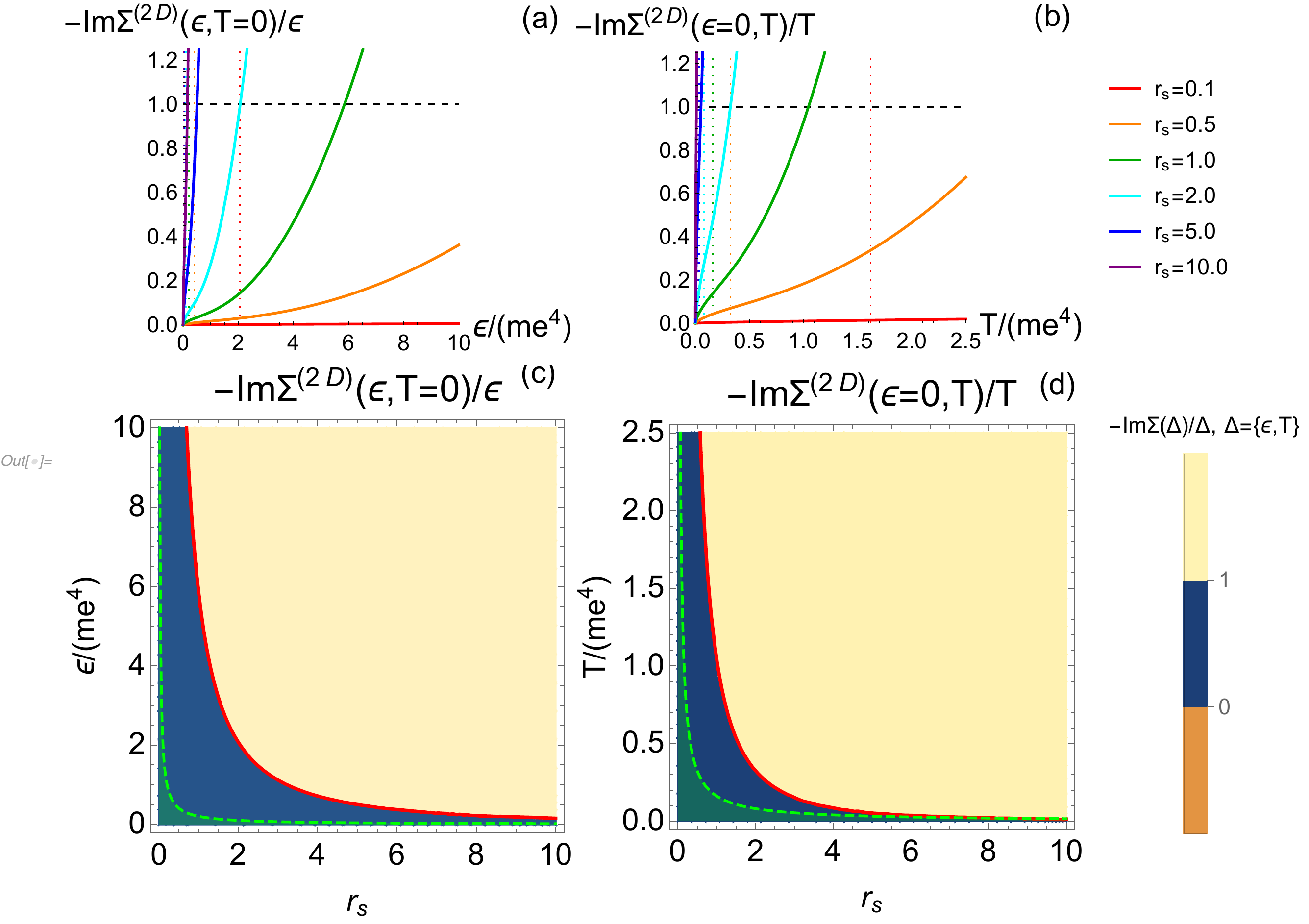}
	\caption{The same as Figs.~\ref{fig:p-3}(a)-(b)  and~\ref{fig:cp-3}(a)-(b)  but plotted in fixed units.
	Note that the nonmonotonic $r_s$ dependence in Fig.~\ref{fig:p-3}(a)-(b) disappears after using the fixed units.}
	\label{fig:F}
\end{figure}

One compelling conclusion of the results presented in Figs.~2 -7 is that the Fermi liquid theory and the quasiparticle picture are extremely robust in both 2D and 3D interacting systems, generically remaining valid at least up to an energy $\Ef$ above the Fermi level.  In general, the regime of this robustness decreases with increasing $r_s$, but even for $r_s \sim 10$, where our RPA theory is  suspect, the quasiparticles remain well-defined (i.e., the magnitude of $\im \Sigma^{(R)}$ is less than $\e$ and/or $T$) up to an energy $\Ef$ above the Fermi level.  The stability is quantitatively slightly weaker in 2D than in 3D, but the difference is not significant enough to draw any conclusion.

\subsection{Effective mass}\label{sec:mass}

As mentioned earlier, in the Fermi liquid theory, an interacting electron system is composed of quasiparticles whose effective mass is different from the bare mass of electron but is renormalized by the electron-electron interactions.
The effective mass of the quasiparticle is a fundamental parameter in the Fermi liquid theory, and has been extensively studied before. See for example, Refs.~\cite{Galitskii1958,Rice,Gell-Mann,Vinter,DS,Galitski2004,Zhang} for electron systems with Coulomb interactions and Refs.~\cite{Chubukov2003,Chubukov2004} for the case of the short-range interactions. 
In this section, we use the analytical expressions for the real part of the self-energy provided in Sec.~\ref{sec:theory} to rederive the effective mass for Coulomb interactions. We also provide a higher order $T^2$ correction to the previous result~\cite{Galitski2004} in both 2D and 3D.

To the leading order in dynamically screened interactions, the effective mass can be obtained from the real part of the self-energy using the following formula~\cite{Galitski2004}:
\begin{align}
\begin{aligned}
	\frac{m^*(T)}{m}
	=
	\dfrac
	{1-\frac{\partial}{\partial \e } \re \Sigma^{(R)} (\e,\xi_{\kb}) }
	{1+\frac{\partial}{\partial \xi_{\kb} }  \re \Sigma^{(R)} (\e,\xi_{\kb}) }
	\bigg|_{\e=\xi_{\kb}=0}
	\approx
	1
	-
	\left( 
	\frac{\partial}{\partial \e} 
	+
	\frac{\partial}{\partial \xi_{\kb}}
	\right)  
	\re \Sigma^{(R)} (\e,\xi_{\kb}) \bigg|_{\e=\xi_{\kb}=0}.
\end{aligned}
\end{align}
Here $m^*$ and $m$ denote the effective mass and bare mass, respectively.

Inserting the low energy limit ($\e\ll T$) expression for the real part of the on-shell  ($\e=\xi_{\kb}$) self-energy (Eq.~\ref{eq:ReSig-<}) into the equation above, we obtain the effective mass for 2D systems
\begin{align}\label{eq:2D-m}
\begin{aligned}
	\frac{m^*(T)}{m}
	=\,&
	1
	-
	\frac{ r_s}{\sqrt{2}\pi}
	\ln \left( \frac{2\sqrt{2}}{r_s} \right) 
	+
	\frac{ \ln 2}{4} 
	\frac{T}{\Tf}
	-
	\frac{5\pi}{48\sqrt{2} r_s}
	\frac{ T^2 }{\Tf^2} 
	\ln \left( \frac{r_s \Tf}{ T }\right)
	\\
	&
	-
	\left[
	-
	\frac{\pi}{96 \sqrt{2} }
	\left(32-10\gamma_E-25\ln 2 \right) 
	-
	\frac{5}{8 \sqrt{2} \pi }
	\left( \zeta ' (2) + \frac{\pi^2}{6}\ln 2\right) 
	\right] 
	\frac{ T^2 }{\Tf^2 r_s} .
\end{aligned}
\end{align}
This expression is valid for quasiparticles with $\e \ll T$  in the regime of $r_s \ll 1$ and $r_s^{3/2} \ll T/ \Ef \ll r_s$.
Similarly, applying Eq.~\ref{eq:3D-ReSig-<}, we find that the 3D effective mass takes the following form for $r_s\ll {T}/{\Tf} \ll \sqrt{r_s}$:
\begin{align}
\begin{aligned}\label{eq:3D-m}
	\frac{m^*(T)}{m}
	=\,&
	1
	-
	\frac{1}{3\pi^{4/3}} (\frac{2}{3})^{2/3} r_s
	\ln \left( \frac{3\pi^2}{2r_s^{3/2}}\right) 
		+
	\frac{4-\pi^2}{192}	\pi^2 
	\frac{T^2}{ \Ef^2} 
	\ln \left( \frac{1}{2(\frac{16}{3\pi^2})^{1/3} \sqrt{r_s}} \frac{T}{\Ef}\right) 	
	\\
	&
	+
	\left[ 
	\frac{4-\pi^2}{384}
	\left( 3-2\gamma_E-2\ln 2 \right) 
	-
	\frac{2}{3} C_1
	+
	\frac{4-\pi^2}{32\pi^2}	
	\left( \frac{\pi^2}{6} \ln 2+\zeta'(2)\right) 
	\right] 
	\pi^2
	\frac{T^2}{ \Ef^2}.
\end{aligned}
\end{align}
We note that the applicable temperature range of this result is different from the one in Ref.~\cite{Galitski2004}.
Our current results include both leading and subleading temperature corrections to the renormalized quasiparticle effective mass in both 2D and 3D.

We emphasize that, in 2D, the leading order temperature correction to the effective mass is linear in $T$ and  independent of $r_s$~\cite{DS,Galitski2004}.
This linear-in-$T$ correction has also been found for 2D systems with short-range interactions~\cite{Chubukov2003,Chubukov2004}.
By contrast, for 3D systems, the leading order temperature correction is of the order of $T^2\ln T$, much smaller compared with its 2D counterpart. The corresponding coefficient is also $r_s$-independent. 
The appearance of the linear-in-$T$ leading order effective mass renormalization in 2D compared with the leading order $T^2 \ln T$ renormalization in 3D implies much stronger interaction effects in 2D compared with 3D.
But, this stronger effective 2D interaction does not imply any failure of the Fermi liquid theory.

\subsection{Hydrodynamic and ballistic regimes}\label{sec:hydro}


Hydrodynamics is a useful approach to describe physical processes of interacting systems at time and length scales that are much larger compared with the ones associated with local equilibration, provided that the dominant microscopic inter-particle scattering process is momentum-conserving.
Electron liquids can be described hydrodynamically if the dominant scattering process is the electron-electron collision which is responsible for local equilibration. 
In this case, the electron system reaches local equilibrium at the time scale of the electron-electron inelastic scattering time, which is much shorter than the other time scales in the problem, and therefore obeys hydrodynamics. 
By contrast, if electron-impurity and/or electron-phonon scattering is stronger than the electron-electron scattering so that momentum conservation is not preserved, hydrodynamics does not apply.  We assume here that the electron-impurity and electron-phonon scattering are negligible in the system under consideration.

The main condition for hydrodynamics to be applicable to an electron liquid~\cite{pines,Lucas,Jian}  is that the electron-electron scattering time $\tau_{\e}$  is much smaller than the time scale or inverse frequency of the physical process being studied ($\e^{-1}$):
\begin{align}\label{eq:hydro}
\begin{aligned}
\tau_{\e} \ll \e^{-1}.
\end{aligned}
\end{align}
Equivalently, the mean free path  $l=\vf \tau_{\e}$ associated with electron-electron collisions should be much smaller compared with the external length scale $q^{-1}$:
\begin{align}
	 l \ll q^{-1}.
\end{align}
Here energy $\e$ can be considered as an external frequency for an experiment probing the response of the system to an applied field, and $q$ is the corresponding external wavevector which is related to external energy $\e$ by $\e=q\vf$.
The collision-dominated regime where Eq.~\ref{eq:hydro} is satisfied is known as the hydrodynamic regime. On the other hand, the collisionless regime where the hydrodynamics condition Eq.~\ref{eq:hydro} no longer applies is usually called the  ballistic regime.

As mentioned earlier, the electron-electron inelastic scattering rate can be extracted from the imaginary part of the self-energy:
\begin{align}\label{eq:tau}
\begin{aligned}
	\tau^{-1}_{\e}=-2\im \Sigma^{(R)} (\e, T).
\end{aligned}
\end{align}
We then insert the previously obtained formulas for $\im \Sigma^{(R)} (\e, T)$ for arbitrary $\e/T$  to investigate the hydrodynamic and ballistic regimes.
As in Sec.~\ref{sec:applicability}, we assume the self-energy formulas are applicable even for low densities,  high temperatures and high energies.

\begin{figure}[t!]
	\centering
	\includegraphics[width=0.7\linewidth]{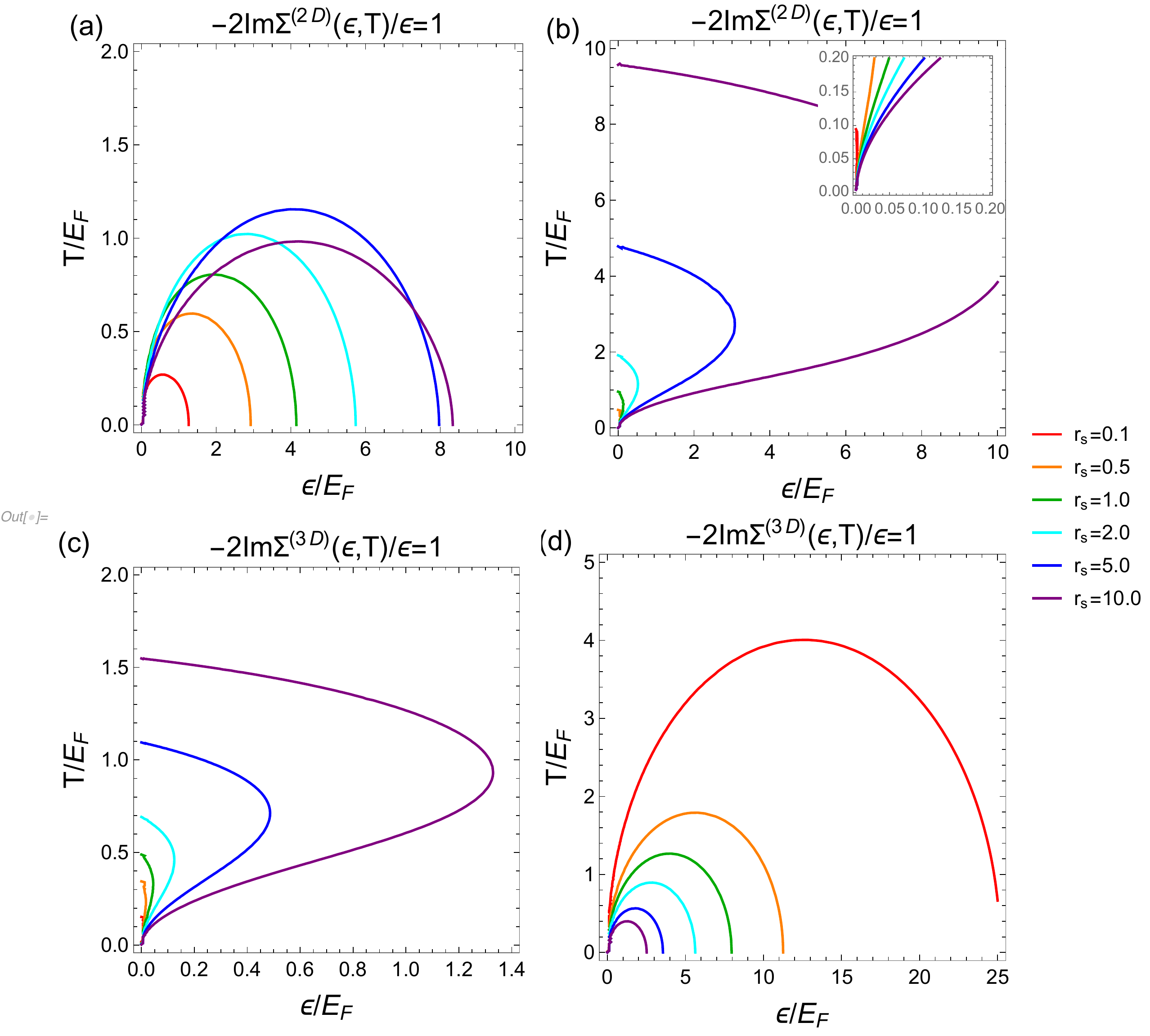}
	\caption{Boundaries separating the hydrodynamic regimes where $-2\im \Sigma^{(R)} (\e, T)/\e>1$ and the ballistic  regimes where $-2\im \Sigma^{(R)} (\e, T)/\e<1$ for 2D (upper panels) and 3D (lower panels) electron systems for $r_s=0.1, 0.5, 1.0, 2.0, 5.0, 10.0$.
	The left panels are obtained using all terms in the self-energy expressions Eqs.~\ref{eq:ImSig} and~\ref{eq:3DimSig} (which are applicable to arbitrary $\e/T$), while for the right panels, the highest order terms in these equations are neglected.
	In the $(\e/\Ef, T/\Ef)$-plane, the hydrodynamics regimes lie in the low energy (high temperature) part, whereas the ballistic  regime correspond to the high energy (low temperature) part.
	The inset of panel (b) zooms in the lower-left (low energy and temperature) part to clearly show the behavior of $r_s=0.1$ case. 	
}
	\label{fig:H}
\end{figure}

Using Eq.~\ref{eq:hydro} and Eq.~\ref{eq:tau}, it is straightforward to see that the hydrodynamic regime should follow $-2\im \Sigma^{(R)} (\e, T)/\e>1$, whereas the ballistic  regime satisfies $-2\im \Sigma^{(R)} (\e, T)/\e<1$.
In Fig.~\ref{fig:H}, in the $(\e/\Ef, T/\Ef)$-plane, we depict contours given by the following condition for various $r_s$
\begin{align}
\begin{aligned}
		-2\im \Sigma^{(R)} (\e, T)/\e=1.
\end{aligned}
\end{align}
These contours divide the plane into two parts: the low energy (high temperature) part corresponds to the hydrodynamic regime, while the high energy (low temperature) part is associated with the ballistic  regime.
We consider both 2D and 3D cases which are shown in the upper and lower panels of Fig.~\ref{fig:H}, respectively.
For Fig.~\ref{fig:H}(a) (Fig.~\ref{fig:H}(c)), all terms in the 2D (3D) self-energy expression Eq.~\ref{eq:ImSig} (Eq.~\ref{eq:3DimSig}) are used, whereas for Fig.~\ref{fig:H}(b) (Fig.~\ref{fig:H}(d)), the highest order term in that equation is ignored.
We note that the boundaries of the hydrodynamic and ballistic regimes obtained from different self-energy approximations are quite distinct from each other.
Both approximations are in fact only applicable to sufficiently low energies and temperatures, and the results for $\e/\Ef>1$ or $T/\Ef>1$ in Fig.~\ref{fig:H} are therefore not reliable. 
 
We emphasize that if electron-impurity and/or electron-phonon scattering effects are present, then one must also compare the electron-electron scattering rates with these other scattering rates, and hydrodynamics would apply only if electron-electron scattering is the dominant scattering process.  In real materials it is a challenge to create conditions for the hydrodynamic regime since electron-impurity and electron-phonon scattering tend to dominate at low and high temperatures, respectively.

\section{Wiedemann-Franz (WF) and Kadowaki-Woods (KW) relations for 2D interacting systems}

Given our analytical results for the 2D inelastic scattering rates for electron-electron interactions, we can briefly comment on the repercussions for the well-known Wiedemann-Franz (WF) and Kadowaki-Woods (KW)  relations in 2D Fermi liquids.

The WF law~\cite{WF} states that the ratio $L=\kappa/(\sigma T)$, where $\sigma$ $(\kappa)$ is the electronic electrical (thermal) conductivity, is universal in metals (i.e. Fermi liquids), and the KW law~\cite{KW} states that the ratio $K= \mathcal{A}/\gamma^2$ is universal, where $\mathcal{A}$ $(\gamma)$ is the coefficient of the $T^2$ term (the linear $T$ term) in the temperature dependence of the resistivity (specific heat).  Both of these universalities are obeyed rather widely in metals (for the KW law, the metal must show a dominant $T^2$ dependence in its resistivity limiting it to the so-called strongly correlated materials such as transition metals, heavy fermion compounds and metallic oxides).

Both of these quantities involve the electronic dc conductivity $(\sigma)$ or the resistivity ($\rho=1/\sigma$), which is operationally defined by the following Drude formula:
\begin{align}
	\sigma=\frac{n e^2 \tau}{m^*}.
\end{align}
Here, $n$ is the 2D carrier density and $m^*$ is the effective mass, and the key quantity is the transport relaxation time $\tau$ which depends on the details of the scattering process.  In ordinary metals and doped semiconductors, transport is mostly limited by electron-impurity (at lower temperatures) and electron-phonon (at higher temperatures) scattering of the carriers.  This is consistent with the observed temperature-independence (arising from impurity scattering) and linear-in-$T$ temperature dependence (arising from phonon scattering in the equipartition regime) of the electrical resistivity in simple metals and doped semiconductors at low and high temperatures, respectively.  Consideration of electron-impurity and electron-phonon interactions is beyond the scope of the current work - see, e.g., Refs.~\cite{resistivity-1,resistivity-2,resistivity-3}. 

Our focus in the current work is electron-electron interaction, which, by virtue of the Galilean invariance in a continuum system, cannot affect the resistivity since electron-electron scattering usually conserves total momentum of the system (and transport is a momentum relaxation phenomenon).  It is, however, often claimed in the literature that a hallmark of a Fermi liquid is the manifestation of a dominant $T^2$ temperature dependence of the resistivity, with the temperature-dependent part of the resistivity going primarily as
\begin{align}
	\rho(T) =1/\sigma (T)=\mathcal{A} T^2.
\end{align}
Although no simple metal or 2D doped semiconductor shows such a $T^2$ temperature dependence in the resistivity, many strongly correlated metals (e.g. transition metals, heavy fermion compounds, various metallic oxides) do.  Since the intraband electron-electron scattering cannot contribute to the resistivity by virtue of its explicit momentum conserving nature, the widespread assumption is that the $T^2$ resistivity manifesting in many narrow band metals arises from electron-electron scattering involving umklapp or interband scattering processes in a Fermi liquid which do not conserve total momentum.  We accept this assumption uncritically, and consider the consequences for such a $T^2$ resistivity arising from electron-electron scattering for the WF and KW relations using our theory for the inelastic electron-electron interaction induced scattering rate.  

This will involve the assumption that our calculated imaginary part of the on-shell electron self-energy at the Fermi energy is indeed the appropriate scattering for the resistivity so that we can make the identification that our calculated imaginary part of the temperature-dependent electron self-energy at the Fermi energy defines the transport scattering relaxation time $\tau$ entering the electrical resistivity through:
\begin{align}
	1/\tau = -2 \im \Sigma^{(R)} (0, T) .
\end{align}
Using our analytical results for $\im \Sigma^{(R)}$ (Eq.~\ref{eq:ImSig-<}), we get:
\begin{align}
	\frac{1}{2 \tau}
	=	
	\frac{\pi}{8} 
	\frac{ T^2 }{\Ef} 
	\ln \left( \frac{\sqrt{2} r_s \Ef}{T}\right) 
	-
	\frac{\pi}{24}
	\left(6 + \ln 2\pi^3 -36 \ln \mathrm{A} \right) 
	\frac{ T^2}{\Ef}
	+ 
	\frac{7\zeta(3)}{2\sqrt{2}\pi} \frac{T^3}{r_s\Ef^2}.
\end{align}
With this expression for $\tau$ along with the Drude formula for the dc resistivity, we now discuss below the WF and KW relations individually for 2D systems.

\subsection{WF Law}

For the WF law, which is a statement on the ratio of the electrical and thermal conductivity, we can write~\cite{WF-1} 
\begin{align}
	L=\frac{\kappa}{\sigma T} = \frac{\pi^2}{3}   \frac{\tau}{\tau + \tau_i}.
\end{align}
Here, $\tau_i$ is the temperature-independent momentum relaxation time due to electron-impurity elastic scattering which must dominate at $T=0$, where $\tau (T)$ above, arising from inelastic electron-electron scattering, becomes infinite.  Since $1/\tau \sim -T^2 \ln (T/T_F)$, we conclude that the relevant WF law for 2D strongly correlated systems becomes:
\begin{align}
	L \sim  \dfrac{L_0 }{ 1 + B (T/T_F)^2 \ln (T/T_F)}.
\end{align}
Here the nonuniversal materials and sample dependent constant $B$~\cite{Catelani} depends both on the strength of the impurity scattering and the prefactor of the $T^2 \ln T$ term in our imaginary self-energy, and $L_0$ is the ideal Lorenz number for the WF law which must be recovered at $T=0$ where the electron-electron scattering vanishes.  The important qualitative result, however, is that the WF law will be strongly violated if the electron-electron scattering induced inelastic scattering is strong in the system, and at low temperatures this violation, or equivalently the suppression of the Lorenz number from its ideal WF value, will follow a $[1 + B (T/T_F)^2 \ln (T/T_F)]^{-1}$ dependence in a Fermi liquid 2D metal (with $B$ being a materials-dependent nonuniversal number).  This violation of the WF law in 2D metals should be observable for strongly interacting systems at low enough temperatures, where the electron-phonon interaction is negligible (but electron-electron interaction is still significant).  Note that we can further refine the violation of the WF law by adding the next-to-leading-order temperature dependent terms in the imaginary part of the self-energy  (Eq.~\ref{eq:ImSig-<}).
We note that the $\ln T$ term in above 2D equations is most likely absent in the transport coefficients, but this is beyond the scope of the current work.

\subsection{KW law}

For the KW law, we need an expression for the 2D electronic specific heat, which in the leading-order Fermi liquid theory can be written as:
\begin{align}
	C_e = C (m^* T),
\end{align}
where $C$ depends only on universal constants, and $m^*$ is the carrier effective mass.  Thus, $\gamma=Cm^*$ for the 2D Fermi liquid. Note that in 2D, carrier density or Fermi energy does not enter the expression for the electronic specific heat by virtue of the constancy of the 2D density of states.  For the KW relation, we need the ratio $K$ of the $T^2$ term in the resistivity with the linear-in-$T$ term in the specific heat, which is given by:
\begin{align}
K= \mathcal{A}/(Cm^*)^2 
\end{align}
with $\mathcal{A}$ given by:
\begin{align}
\mathcal{A}= \frac{m^*}{ne^2} \frac{1}{\tau}\frac{1}{T^2} = \mathcal{A}' \frac{m^*}{n}  \frac{1}{E_F} \ln \left( \frac{T_F}{T}\right) ,
\end{align}
where $\mathcal{A}'$ depends on universal constants.  Remembering that in 2D Fermi systems, $\Ef \sim (n/m^*)$, we get, combining everything, for the KW ratio:
\begin{align}
	K=\frac{B' }{n^2}\ln  \left( \frac{T_F}{T}\right) ,
\end{align}
where $B'$ is universal within the free electron type Fermi liquid theories. (Again, the $\ln T$ is most likely absent, but this is beyond the scope of the current work.) We note that within the free electron type band structure model for Fermi liquids, the Kadowaki-Woods ratio goes as $n^{-2}$ in 2D in contrast to the corresponding $n^{-7/3}$ dependence in 3D systems.  This precise difference by a factor of $n^{-1/3}$ can be understood as arising from the dimensional difference between 2D and 3D resistivity definitions: in 2D the resistivity is simply measured in ohms whereas in 3D it is measured in ohm.cm, so the two units must differ by a unit of length, which is precisely what $n^{-1/3}$ is for a 3D carrier density $n$.    It may be worthwhile to emphasize that the KW relation is by no means universal either in 2D or in 3D by virtue of the strong density dependence inherent in the KW ratio.  It is only when different materials have similar effective carrier densities, one can talk about a universal KW relation, and even then it is rather a dubious universality since the origin of the $T^2$ resistivity term may differ from system to system.

\section{Conclusion}

We have analytically investigated the domain of validity of the Fermi liquid theory and the quasiparticle picture in Coulomb interacting continuum 2D and 3D electron liquids.  Using the leading-order dynamical screening approximation, which is exact in the high-density limit, we calculate exact expressions for the real and imaginary parts of the electron self-energy in expansions in temperature and energy measured from the Fermi surface.  
Using the calculated imaginary part of the self-energy, we estimate the $r_s$-dependent crossover energy and temperature scales above which well-defined quasiparticles no longer exist.  We find that in both 2D and 3D systems the quasiparticle picture remains valid at energies and temperatures comparable to Fermi energies or above, implying a robust validity of Fermi liquid theory, not only in 3D, but also in 2D.  In general, the energy scales for the validity of the Fermi liquid theory decreases with increasing $r_s$, but the theory being exact only for small $r_s$, this finding, although intuitively appealing, is only tentative.  Recent work, which calculates both the on-shell and the off-shell self-energy numerically without making $\varepsilon\ll \Ef$ approximation, has also concluded that the regime of validity of the 2D Fermi liquid theory and quasiparticle concept is wide and applies to very high energies~\cite{Ahn}.

We emphasize that the applicability of the theory for $r_s>1$ is irrelevant to our work since we are theoretically addressing a matter of principle:  how high in $\e$ ($T$) does the theory still remain valid? All we need is the controlled nature of the theory, which is true for $r_s\ll1$, in the weak coupling perturbative RG sense.
We mention that RPA itself is a controlled approximation giving exact results for small $r_s$ although it is known to work very well empirically at metallic densities (with $r_s \sim 4-6$).  The reason is a likely cancellation of higher-order diagrams, but no systematic theory exists for $r_s>1$ because of the failure  of the $r_s$-expansion.  Quantum Monte Carlo calculations can be used in some situations to study the ground state properties, but are inapplicable to the question of the validity of the Fermi liquid theory (we are asking in this work) which must study the quasiparticle excitations that cannot be done by Monte Carlo techniques.  What we show here is that the Fermi liquid theory applies at very high energies away from the Fermi surface as well as at very high temperatures not only at small $r_s$, but also at intermediate to large $r_s$.  Our results for $r_s>1$ are obviously not exact, but as long as there is no ground state strong-coupling quantum phase transition, the theory should remain valid qualitatively at higher $r_s$.

We also provide 2D and 3D `phase diagrams' for temperature and frequency, where the electron liquid crosses over from the collisionless ballistic regime to the collision dominated hydrodynamic regime neglecting effects of impurity and phonon scattering, and obtained general expressions for 2D Wiedemann-Franz and Kadowaki-Woods relations, using our calculated self-energy expressions.  We have also provided effective mass renormalization results for 2D and 3D systems including leading and subleading temperature corrections, extending existing results in the literature.

\section*{Acknowledgement}

This work is supported by the Laboratory for Physical Sciences (SDS) and the 
Simons Foundation “Ultra-Quantum Matter” Research 
Collaboration (YL).

\begin{appendices}

\appendixheaderon

\section{Derivation of self-energy formulas using Matsubara technique}\label{sec:matsubara}

For the sake of completeness, in this Appendix we briefly review an alternative derivation of the self-energy formula Eq.~\ref{eq:Sig1} using the Matsubara instead of Keldysh technique. For more details, see for example Refs.~\cite{AGD,fetter}.

As explained in earlier sections, for electrons interacting via Coulomb interactions,  in the high density limit $r_s \ll 1$, the electron self-energy is given by the RPA self-energy represented by the diagram in Fig.~\ref{fig:D}(b).
In the Matsubara formalism, the black sold line represents the noninteracting Matsubara Green's function for electrons, which acquires the form
\begin{align}	\label{eq:GR-m}
	&	G_0 (\kb,i\e_n) = \left( i\e_n - \xi_{\kb}  \right)^{-1}.
\end{align}		
 Here $\e_n=2\pi (n+1/2)T$ is the fermionic Matsubara frequency.
The red wavy line with a solid dot stands for the dynamically screened RPA interaction $D$ defined diagrammatically by the Dyson equation shown in Fig.~\ref{fig:D}(a).  In particular, $D$ is related to the bare interaction $V$ (red wavy line with a open dot) and the polarization bubble $\Pi$ (black bubble) through the following equation:
\begin{align}
\begin{aligned}
D(\qb,i\ww_m)
=
\left[ V^{-1}(\qb)-\Pi(\qb,i\ww_m) \right]^{-1},
\end{aligned}
\end{align}
with $\ww_m=2\pi mT$ being the bosonic Matsubara frequency.
 $\Pi(\qb,i\ww_m)$ is given by a product of two bare electron Green's functions:
\begin{align}\label{eq:M-Pi}
\begin{aligned}
	\Pi(\qb,i\ww_m)
	=
	2T\sum_{\e_n}
	\int \frac{d^d \kb}{(2\pi)^d}
	G_0(\kb+\qb,i\e_n+i\ww_m)
	G_0(\kb,i\e_n),
\end{aligned}
\end{align} 
where the overall factor of two arises from the summation over spin indices.
It is then straightforward to see that the RPA self-energy depicted in Fig.~\ref{fig:D}(b) evaluates to
\begin{align}
\begin{aligned}
	\Sigma(\kb,i\e_n)
	=
	-T \sum_{\ww_m}
	\int \frac{d^d \qb}{(2\pi)^d}
	D(\qb,i\ww_m)
	G_0(\kb+\qb,i\e_n+i\ww_m).
\end{aligned}
\end{align}

After analytical continuation, the equation above then leads to the retarded self-energy
\begin{align}
\begin{aligned}
	\Sigma^{(R)}(\kb,\e)
	=
	\frac{i}{2}
	\int \frac{d^d \qb}{(2\pi)^d}
	\int \frac{d \ww}{2\pi}
	&
	\left\lbrace 
	\left[ 
	D^{(R)}(\qb,\ww)
	-
	D^{(A)}(\qb,\ww)
	\right] 
	G_0^{(R)}(\kb+\qb,\e+\ww)
	\coth\left( \frac{\ww}{2T}\right) 
	\right. 
	\\
	&\left. 
	+
	D^{(A)}(\qb,\ww)
	\left[ 
	G_0^{(R)}(\kb+\qb,\e+\ww)
	-
	G_0^{(A)}(\kb+\qb,\e+\ww)
	\right] 
	\tanh\left( \frac{\ww+\e}{2T}\right) 
	\right\rbrace.
\end{aligned}
\end{align}
With the help of the FDT Eqs.~\ref{eq:FDT-G} and~\ref{eq:FDT-D}, one can prove that this equation is equivalent to the self-energy formula Eq.~\ref{eq:Sig1} derived in the Keldysh formalism.
Similarly, by analytical continuation, Eq.~\ref{eq:M-Pi} can be transformed to
\begin{align}
\begin{aligned}
	\Pi^{(R)}(\qb,\ww)
	=
	-i
	\int \frac{d^d \kb}{(2\pi)^d}
	\int \frac{d \e}{2\pi}
	&
	\left\lbrace 
	G_0^{(R)}(\kb+\qb,\e+\ww)
	\left[ G_0^{(R)}(\kb,\e)-G_0^{(A)}(\kb,\e)\right] 
	\tanh\left( \frac{\e}{2T}\right) 
	\right. 
	\\
	&\left. +
	\left[ 
	G_0^{(R)}(\kb+\qb,\e+\ww)
	-
	G_0^{(A)}(\kb+\qb,\e+\ww)
	\right] 
	G_0^{(A)}(\kb,\e)
	\tanh\left( \frac{\e+\ww}{2T}\right) 
	\right\rbrace ,
\end{aligned}
\end{align} 
which is equivalent to the retarded polarization operator formula Eq.~\ref{eq:Pi2} derived in Sec.~\ref{sec:formula}.
	
\section{Integrals involving hyperbolic functions}\label{sec:integrals}

In this Appendix, we provide analytical results for integrals of the form $I_{1,2}(a)$ in Eq.~\ref{eq:I12},
which are needed for the calculation of electron self-energy with arbitrary value of $\e/T$.
The results are obtained with the help of the exponential series expansion of the hyperbolic functions (Eq.~\ref{eq:expser}). In particular, $I_{1,2}(a)$ can be rewritten as
\begin{subequations}\label{eq:I12-s}
\begin{align}
	&\begin{aligned}\label{eq:cothf}
	 I_1(a)
	=&
	2\sum_{k=1}^{\infty}  
	\left[ 
	2
	-
	(-1)^ke^{-2ka}
	-
	(-1)^k e^{2ka}
	\right] 
	\int_0^{\infty} dx f(x)e^{-2kx} 
	\\
	&+
	4\sum_{k=1}^{\infty}  
	\int_{0}^{a} dx 
	f(x)
	(-1)^k
	\cosh \left( {2kx-2ka} \right) 
	+
	2
	\int_{0}^{a}
	dx
	f(x),
	\end{aligned}
	\\
	&\begin{aligned}\label{eq:tanhf}
	I_2(a)
	=\,&
	2\sum_{k=1}^{\infty} 
		(-1)^k
	\left[ 
	e^{-2ka}
	-
	e^{2ka}
	\right]
	\int_0^{\infty} dx f(x)e^{-2kx} 
	\\
	&+
	4 \sum_{k=1}^{\infty} 
	\int_0^{a} dx f(x)
	(-1)^k
	\cosh\left( {2kx-2ka}\right) 
	+
	2 \int_0^{a} dx f(x).
\end{aligned}
\end{align}
\end{subequations}
For both $I_1(a)$ and $I_2(a)$, the last two terms cancel with each other due to the fact that $\sum_{k=1}^{\infty} (-1)^k \cosh(2kx-2ka)=-1/2$, and we are left with the first terms in both Eq.~\ref{eq:cothf} and Eq.~\ref{eq:tanhf}.
By inserting the result of the integral $ \int_0^{\infty} dx f(x)e^{-2kx} $ into Eq.~\ref{eq:I12-s}, one is able to evaluate $I_{1,2}(a)$ for various $f(x)$:
\allowdisplaybreaks
\begin{align}\label{eq:hy}
\begin{aligned}
&
\int_0^{\infty} d x
\left[ 2 \coth\left( x\right) 
-\tanh\left( x+a\right)-\tanh\left( x-a \right) \right] 
x
=\,
\frac{\pi^2}{4}+a^2,
\\
&\int_0^{\infty} d x
\left[ 2 \coth\left( x\right) 
-\tanh\left( x+a\right)-\tanh\left( x-a \right) \right] 
x\ln x
\\
&=\,
\left( 1- \gamma_E-\ln 2 \right) a^2
+\frac{\pi^2}{12}
\left( 3-\gamma_E-\ln \frac{2}{\pi^2} -24 \ln \mathrm{A} \right)
-\frac{1}{2}
\left[ \partial_s \Li_s (-e^{-2a})+ \partial_s \Li_s (-e^{2a})\right] \bigg\lvert_{s=2},
\\
&\int_0^{\infty} d x
\left[ 2 \coth\left( x\right) 
-\tanh\left( x+a\right)-\tanh\left( x-a \right) \right] 
x^2
=\,
\zeta(3)-\frac{1}{2}\left[ \Li_3(-e^{-2a})+  \Li_3(-e^{2a}) \right],
\\
&	\int_0^{\infty}
dx
\left[ 
\tanh (x+a)
-\tanh (x-a)
\right] 
=\,
2a,
\\
&	\int_0^{\infty}
dx
\left[ 
\tanh (x+a)
-\tanh (x-a)
\right] 
x
=\,
\frac{1}{2}\left[ \Li_2(-e^{-2a})-\Li_2(-e^{2a}) \right] ,
\\
&
\int_0^{\infty}
dx
\left[ 
\tanh (x+a)
-\tanh (x-a)
\right] 
x^2
=\,
\frac{2}{3}a^3+\frac{\pi^2}{6}a,
\\
&
\int_0^{\infty}
dx
\left[ 
\tanh (x+a)
-\tanh (x-a)
\right] 
x^2\ln x
=\,
\left( 3-2\gamma_E-\ln 4 \right) 
\left( \frac{1}{3}a^3+\frac{\pi^2}{12} a \right)
+
\frac{1}{2}
\left[ \partial_s \Li_s (-e^{-2a}) - \partial_s \Li_s (-e^{2a}) \right] \bigg\lvert_{s=3}.
\end{aligned}
\end{align}
\end{appendices}

\bibliography{2DFL}

\end{document}